\documentclass[%
 reprint,
superscriptaddress,
 amsmath,amssymb,
 aps,showpacs,longbibliography,footinbib,
pra,
]{revtex4-2}
\usepackage{xcolor}
\usepackage{float}
\definecolor{antiquefuchsia}{rgb}{0.57, 0.36, 0.51}
\definecolor{cadetblue}{rgb}{0., 0.62, 0.63}
\definecolor{brightmaroon}{rgb}{0.76, 0.13, 0.28}

\usepackage{amssymb}
\usepackage{academicons}
\usepackage{amsmath}
\usepackage{amsfonts}
\usepackage{dcolumn}
\usepackage{bm}
\usepackage{mathrsfs}
\usepackage{times}
\usepackage{natbib}
\usepackage{graphicx} 
\usepackage[linesnumbered,ruled,vlined]{algorithm2e}
\usepackage{lipsum}%
\usepackage[percent]{overpic}
\usepackage{scalerel}
\usepackage{tikz}
\usetikzlibrary{svg.path}

\definecolor{orcidlogocol}{HTML}{A6CE39}
\tikzset{
  orcidlogo/.pic={
    \fill[orcidlogocol] svg{M256,128c0,70.7-57.3,128-128,128C57.3,256,0,198.7,0,128C0,57.3,57.3,0,128,0C198.7,0,256,57.3,256,128z};
    \fill[white] svg{M86.3,186.2H70.9V79.1h15.4v48.4V186.2z}
                 svg{M108.9,79.1h41.6c39.6,0,57,28.3,57,53.6c0,27.5-21.5,53.6-56.8,53.6h-41.8V79.1z M124.3,172.4h24.5c34.9,0,42.9-26.5,42.9-39.7c0-21.5-13.7-39.7-43.7-39.7h-23.7V172.4z}
                 svg{M88.7,56.8c0,5.5-4.5,10.1-10.1,10.1c-5.6,0-10.1-4.6-10.1-10.1c0-5.6,4.5-10.1,10.1-10.1C84.2,46.7,88.7,51.3,88.7,56.8z};
  }
}

\newcommand\orcidicon[1]{\href{https://orcid.org/#1}{\mbox{\scalerel*{
\begin{tikzpicture}[yscale=-1,transform shape]
\pic{orcidlogo};
\end{tikzpicture}
}{|}}}}

\usepackage{hyperref}
\definecolor{maroon}{rgb}{0.76, 0.0, 0.08}
\definecolor{darkred}{RGB}{135, 33, 9}
\hypersetup{
    colorlinks = true,
    linkbordercolor = {white},
    citecolor = magenta,
    linkcolor = cadetblue,
    menucolor = cadetblue,
    urlcolor = darkred
}
\renewcommand{\phi}{\varphi} 
\renewcommand{\rho}{\varrho} 

\newcommand{\DPF}{\textbf{DPF}}
\newcommand{\gDPF}{\textbf{gDPF}}
\newcommand{\PICE}{\textbf{PICE}}
\definecolor{orcidlogocol}{HTML}{A6CE39}
\definecolor{darkred}{RGB}{135, 33, 9}
\usepackage{xparse}

\newif\iffirstitem
\firstitemtrue

\SetCommentSty{mycommfont}
\newcommand{\nosemic}{\renewcommand{\@endalgocfline}{\relax}}
\newcommand{\dosemic}{\renewcommand{\@endalgocfline}{\algocf@endline}}
\let\oldnl\nl
\newcommand{\nonl}{\renewcommand{\nl}{\let\nl\oldnl}}

\date{\today}

\begin{document} 



\title{Deterministic particle flows for constraining stochastic nonlinear systems}

\author{Dimitra Maoutsa\,\orcidicon{0000-0002-3553-8658}}
\email{dimitra.maoutsa@tu-berlin.de}
\affiliation{%
 Department of Theoretical Computer Science, Technical University of Berlin,
 Marchstraße 23, 10587 Berlin, Germany 
}%
\affiliation{%
 Institute of Mathematics, University of Potsdam,
 Karl-Liebknecht-Str. 24/25, 14476 Potsdam, Germany
}%

\author{Manfred Opper}%
 \email{manfred.opper@tu-berlin.de}
\affiliation{%
 Department of Theoretical Computer Science, Technical University of Berlin,
 Marchstraße 23, 10587 Berlin, Germany\\ 
}%

\affiliation{Centre for Systems Modelling and Quantitative Biomedicine, University of Birmingham, United Kingdom}

\begin{abstract}

Devising optimal interventions for constraining stochastic systems is a challenging endeavour that has to confront the interplay between randomness and nonlinearity. 
Existing methods for identifying the necessary dynamical adjustments resort either to space discretising solutions of ensuing partial differential equations, or to iterative stochastic path sampling schemes. Yet, both approaches become computationally demanding for increasing system dimension.
Here, we propose a generally applicable and practically feasible \emph{non-iterative} methodology for obtaining optimal dynamical interventions for diffusive nonlinear systems.
We estimate the necessary controls from an interacting particle approximation to the logarithmic gradient of two forward probability flows, evolved following deterministic particle dynamics.
Applied to several biologically inspired models, we show that our method provides the necessary optimal controls in settings with terminal-, transient-, or generalised collective-state constraints and arbitrary system dynamics.

\end{abstract}

\maketitle

\section*{Introduction}

Most biological systems are continuously subjected to noise arising either from intrinsic fluctuations due to inherent stochasticities of their constituents, or from external environmental variations at multiple timescales ~\cite{raser2005noise,swain2002intrinsic,hasty2000noise,kepler2001stochasticity}. 
The stochastic nature of these influences confers on these systems complex behaviour~\cite{garcia2012noise,eldar2010functional,simpson2009noise}, but also renders them remarkably unpredictable -  by generating noise induced transitions~\cite{blake2003noise, horsthemke1984noise}, intervening in intracellular communication~\cite{zhou2005molecular}, and compromising precision of biological functions.

Yet, concrete understanding of characteristics, properties, and functions of biological processes often requires external interventions either by precise steering of state trajectories, or by enforcing design constraints that limit their evolution. Characteristic examples of such interventions include modulating transcription pathways to decrease response time, improving stability of epigenetic states in gene regulatory networks~\cite{del2016control}, or modifying cell differentiation in multicellular organisms~\cite{nguyen2021scalable}. 
One then may be interested in statistical properties of constrained trajectories (e.g. for computing averages of macroscopic observables), or in obtaining precise control protocols that implement the imposed limitations. 

In most settings the optimality of the imposed interventions plays critical role. Performing unreasonably strong perturbations may damage the underlying biological tissue, or result in dynamical changes that considerably deviate from physiological biological function. Translated to mathematics this implies the requirement for the interventions to induce the minimum possible deviation from the typical evolution of the unconstrained system.


Such problems can usually be posed as stochastic optimal control problems.
This research area has recently attracted interest in the context of stochastic thermodynamics~\cite{sivak2012thermodynamic,zulkowski2012geometry,gomez2008optimal} and quantum control~\cite{rabitz2000whither,pechen2011there}, i.e. for estimating the free energy differences between two equilibrium states~\cite{hendrix2001fast, shirts2003equilibrium}, or for identifying optimal protocols that drive a system from one equilibrium to another in finite time~\cite{schmiedl2007optimal}. Similar problems appear also often in chemistry, biology, finance, and engineering, required for computation of rare event probabilities~\cite{hartmann2012efficient,chetrite2015variational}, state estimation of partially observed systems~\cite{kim2020optimal,casadiego2018inferring,todorov2009efficient}, or for precise manipulation of stochastic systems to target states~\cite{kappen2005linear,wells2015control} with applications in artificial selection~\cite{nourmohammad2021optimal,evolver}, motor control~\cite{kao2021optimal}, epidemiology, and more~\cite{iolov2014stochastic,scott2004optimal,bernton2019schr,vargas2021solving,exarchos2018stochastic,song2020score}. 
Albeit the prior developments, the problem of controlling nonlinear systems in the presence of random fluctuations remains still considerably challenging.

Central role in (stochastic) optimal control theory plays the Hamilton-Jacobi-Belman (HJB) equation~\cite{bellman1956dynamic}, a nonlinear second order partial differential equation (PDE), characterising the value function of the control problem required for obtaining the optimal controls.
Existing approaches for devising optimal interventions can be broadly divided into two classes: the first class treats the HJB equation directly, while a second class optimises the interventions iteratively by employing stochastic path sampling.
Direct treating the HJB often involves space discretising PDE solvers, that in general scale poorly with system dimension~\cite{garcke2017suboptimal,annunziato2013fokker}.
By introducing certain structural assumptions for the control problem in ~\cite{kappen2005linear}, Kappen proposed the Path Integral (PI-) control formalism that linearises the HJB, and via the Feynman-Kac formula reduces the solution of the stochastic control problem to the computation of a path integral. Thereafter, several methods have either treated the linearised HJB with function approximations~\cite{horowitz2014linear}, or employed path integral approximation methods~\cite{kappen2005path, van2008graphical,rawlik2013path}.
A second class of methods optimises the interventions directly in iterative schemes. A subset of those methods were inspired by the PI-control literature, but instead of focusing on approximating the (exponentiated) value function, they directly optimise the controls by employing information theoretic metrics~\cite{kappen2016adaptive,hartmann2012efficient,zhang2014applications, theodorou2011iterative,thijssen2015path}. 
In particular, the Path integral cross-entropy (PICE) method~\cite{zhang2014applications,kappen2016adaptive},  employs importance sampling to  generate 
paths from a stochastic system with a provisional control and applies appropriate reweighting to iteratevely converge to
the optimal interventions.

\if False

\textcolor{magenta}{  
On the other hand, the Feynman--Kac theorem allows the solution of the linear PDE (Eq.~\eqref{backward_equation}) to be expressed in terms of an integral over
stochastic paths from the uncontrolled system, appropriately reweighted to account for the constraints. 
The {\em path integral cross entropy method} improves over this basic Monte--Carlo approach~\cite{kappen2016adaptive,zhang2014applications} by employing importance sampling. Stochastic paths, generated from an SDE with a non--optimal control, 
are properly reweighted, leading to an iterative optimisation of the control.
uses importance sampling to improve the control iteratively.
Stochastic paths are generated from SDE with a non--optimal, parametric control function.
Path samples are properly reweighted based on the explicit stochastic representation of the optimal path measure $\mathbb{Q}^*$
given by \eqref{Reweight} 
and are used to approximate path averages in the cost function \eqref{KL_formul}.
The optimisation of the approximate cost with respect to the parameters in the control function typically leads to better controls.}

\fi


In this paper, we borrow ideas from the inference formulation of
optimal control~\cite{todorov2008general,kappen2012optimal,levine2018reinforcement,attias2003planning}, and take a new look at the problem by providing a sample based solution that nevertheless avoids stochastic path sampling.
More precisely, we reformulate the optimal controls in terms of the solutions of two
forward (filtering) equations, and employ recent deterministic particle methods for propagating probability flows~\cite{Maoutsa_2020}, properly adapted to fit our purposes.
Building on the theory of time-reversed SDEs~\cite{anderson1982reverse}, we obtain an exact representation of the optimally adjusted drift of the underlying stochastic system in terms of the logarithmic gradient of two forward probability flows.
The latter are 
estimated from an interacting particle approximation to the logarithmic gradient of the density using a variational formulation
developed in the field of machine learning.







We show that we can successfully intervene in a series of biologically inspired systems in time constrained  settings, subjected to terminal, path, and generalised collective state constraints. We further demonstrate how various problem parameters influence the estimated interventions, and compare our framework to the already established Path integral cross entropy method~\cite{kappen2016adaptive}.

\section*{Probability flow optimal control: theoretical background}

\subsection{Constraining stochastic systems with deterministic forcing}\label{Theory:intro}
Biological and physical systems are often subjected to intrinsic or extrinsic noise sources that influence their dynamics. Characteristic examples include molecular reactions and chemical kinetics~\cite{gillespie2003improved}, populations of animal species, biological neurons~\cite{saarinen2008stochastic}, and evolutionary dynamics~\cite{lande2003stochastic,takahata1975effect}.
 Stochastic differential equations (SDEs) effectively capture the phenomenology of the dynamics of such systems
 by both considering deterministic and stochastic forces affecting their state variables $X_t \in  \mathcal{R}^d$ following
\begin{equation}\label{eq:free_SDE}
  dX_t = f(X_t,t) dt  + \sigma dW_t.
\end{equation}
In Eq.~(\ref{eq:free_SDE}) the drift  $f(\cdot,\cdot): \mathcal{R}^d \times \mathcal{R} \rightarrow \mathcal{R}^d$ is a smooth typically nonlinear function that captures the deterministic part of the driving forces,
while $W$ stands for a d--dimensional vector of independent Wiener processes acting as
white noise sources, representing contributions from unaccounted degrees of freedom, thermal fluctuations, or external perturbations. We denote the noise strength by $\sigma \in \mathcal{R}$.
 For the sake of brevity, we consider here additive noise, but the formalism easily generalises for multiplicative and non-isotropic noise, i.e. for a state dependent diffusion matrix $\sigma(x,t)$. In the following we refer to this system as the \emph{uncontrolled} system.

Under multiple independent realisations, the stochastic nature of Eq.~(\ref{eq:free_SDE}) gives rise to an ensemble of trajectories starting from an initial state $X_0=x_0$. This ensemble captures the likely evolution of the considered system at later time points. We may characterise the unfolding of this trajectory ensemble in terms of a probability density $p_t(x)$ for the system state $X_t$, whose evolution is governed by the Fokker--Planck equation
\begin{align}\label{FPE1} 
\frac{\partial p_t(x)}{\partial t} 
&= \nabla\cdot \left[- f(x,t) p_t (x) + \frac{\sigma^2}{2} \nabla p_t(x)\right]\\
&= {\cal{L}}_f p_t(x) , \nonumber
\label{FPoperator}
\end{align}
with initial condition $p_0(x) = \delta(x-x_0)$, and $\mathcal{L}_f$ denoting the Fokker--Planck operator.

Due to the stochastic nature of the system of Eq.~\eqref{eq:free_SDE}, exact pinpointing of its state at some later time point $T$ is in general not possible. 
Yet, often, we desire to drive stochastic systems to predefined target states within a specified time interval.
Characteristic examples include designing artificial selection strategies for population dynamics~\cite{nourmohammad2021optimal}, or triggering phenotype switches during cell fate determination~\cite{wells2015control}. Similar needs for manipulation are also relevant for non-biological, but rather technical systems, e.g. for control of robotic or artificial limbs~\cite{todorov2005stochastic,todorov2004optimality}. In all these settings, external system interventions become essential. 

Here, we are interested in introducing constraints $\mathcal{C}$ to the system of Eq.~(\ref{eq:free_SDE}) acting within a predefined time interval $[0,\; T]$. The set of possible constraints $\mathcal{C}$ comprises terminal $\chi(X_T)$, and path constraints $U(x,t), \text{for } t\leq T $, depending on whether the desired limiting conditions apply for the entire interval, or only at the terminal time point.
The path constraints $U(x,t): \mathcal{R}^{d}\times \mathcal{R} \rightarrow \mathcal{R} $ penalise trajectories (\emph{paths}) to render specific regions of the state space more (un)likely to be visited, while the terminal constraint $\chi(x): \mathcal{R}^{d} \rightarrow \mathcal{R}$ influences the system state $X_T$ at the final time $T$.

To incorporate the constraints $\mathcal{C}$ into the system, we define a modified dynamics, the {\em controlled} dynamics, 
through a change of probability measure of the path ensemble $\mathbb{P}_f$ induced by its uncontrolled counterpart.
More precisely, we define the path measure
$\mathbb{Q}^*$, induced by the controlled system, by a {\em reweighting} of paths
$X_{0:T}$ generated from the uncontrolled one (Eq.~\eqref{eq:free_SDE}) over the time interval $[0,\; T]$ (Supplementary Information). 
Path weights are given by the likelihood ratio ({\em Radon--Nikodym derivative})
\begin{equation}
\frac{d\mathbb{Q}^*}{d\mathbb{P}_f} (X_{0:T}) = \frac{\chi(X_T)}{Z} \exp\left[- \int_0^T U(X_t,t) dt \right],
\label{Reweight}
\end{equation}
where $Z$ is the normalising constant 
\begin{equation}
Z = \Bigg \langle \chi(X_T) \exp\left(- \int_0^T U(X_t,t) dt \right) \Bigg\rangle_{\mathbb{P}_f},
\end{equation}
and $\langle \cdot \rangle_{\mathbb{P}_f}$ denotes the expectation over paths of the uncontrolled system.


By a direct calculation (see Supplementary Information), we show that the infinite dimensional path measure $\mathbb{Q}^*$ is the solution of the variational problem
 \begin{align}
\mathbb{Q}^* = \arg\min_{\mathbb{Q}}\Bigg\{KL(\mathbb{Q} || \mathbb{P}_f)  &+ \int_0^T \Big\langle U(X_t ,t) \Big\rangle_{\mathbb{Q}} \; dt  \nonumber \\ &- \ln \chi(X_T)\Bigg\},
\end{align}
where $KL(\mathbb{Q} || \mathbb{P}_f)$ stands for the relative entropy (\emph{Kullback-Leibler (KL-) divergence}) between the controlled and the uncontrolled path measures.

It can be shown that the optimal path measure $\mathbb{Q}^*$ is induced by a time- and state- dependent perturbation $u(x,t): \mathcal{R}^d \times \mathcal{R} \rightarrow \mathcal{R}^d$ of the deterministic forces $f(x,t)$ acting on the uncontrolled system~\cite{kappen2016adaptive}.  
Thus, we express the controlled dynamics as a space- and time-dependent perturbation of the uncontrolled system
 \begin{align} \label{eq:SDE_controlled}
     dX_t &= \Big( f(X_t,t)   + u(X_t,t) \Big) \; dt + \sigma dW_t \nonumber \\ &= \hspace{25pt}g(X_t,t)\;\hspace{5pt} dt \hspace{30pt}+ \sigma dW_t.
 \end{align}

To identify the optimal interventions we minimise the cost functional $\mathcal{J}$
\begin{equation}
 \mathcal{J} \doteq \min_{u} \Big\langle  \int_{0}^T \left( \frac{1}{2\sigma^2} \|u(x,t)\|^2 + U(x,t)\right)\,  dt -  \ln \chi(X_T) \Big\rangle_{\mathbb{Q}}.
 \label{eq:cost_to_go}
\end{equation}
The first part of the cost functional penalises large interventions $u(x,t)$, and results from minimising the relative entropy between the path measures induced by the controlled and uncontrolled dynamics, $KL(\mathbb{Q} || \mathbb{P}_f)$.
The second term constrains the transient behaviour of the system through the path costs $U(x,t)$, while $\chi(X_T)$, influences only the terminal system state. 

Finding exact optimal 
controls for general stochastic control problems amounts to solving the Hamilton--Jacobi--Bellman (HJB) equation (see~\cite{bellman1956dynamic}), a {\em nonlinear},
partial differential equation (PDE) that is in general computationally demanding to treat directly.

The control cost formulation of Eq.\eqref{eq:cost_to_go} gives rise to a cluster of stochastic control problems known as 
 {\em Kullback-Leibler (KL)-control}~\cite{todorov2007linearly} or \emph{Path integral (PI)-control}~\cite{kappen2005linear,kappen2012optimal} in the literature (see Supplementary Information for details).
For this class of problems  the logarithmic Hopf-Cole transformation~\cite{fleming1977exit}, ie. setting ${\mathcal{J}(x,t) = -\ln( \phi_t(x))}$, linearises the Hamilton-Jacobi Bellman equation~\cite{kappen2005linear}, and the optimally perturbed drift simplifies into (see Supplementary Information) 
\begin{equation} 
g(x,t)  = f(x,t) +  \sigma^2 \nabla \ln \phi_t(x), 
\label{newdrift}
\end{equation}
where the function $\phi_t(x)$ is a solution to the backward PDE 
\begin{equation}
\frac{\partial \phi_t(x)}{\partial t} + {\cal{L}}^{\dagger}_f \phi_t(x) - U(x,t) \phi_t(x) = 0  ,
\qquad
\label{backward_equation}
\end{equation}
with terminal condition $\phi_T(x) = \chi(x) $, and  $\mathcal{L}^{\dagger}_f$ denoting the 
adjoint Fokker--Planck operator. For $U(x,t)\equiv 0$, Eq.\eqref{backward_equation} reduces to the Kolmogorov backward equation for the uncontrolled system.

Although the controlled drift admits a well defined expression in the terms of the solution of the backward partial differential equation of Eq.(\ref{backward_equation}), direct solutions with space discretising schemes~\cite{garcke2017suboptimal, annunziato2013fokker} often suffer from high computational complexity with
increasing dimensionality and become {inefficient} for most practical settings. 
Similarly, stochastic path sampling frameworks building on the the equivalence between path reweighting and optimal control, like the {\em path integral cross entropy method}~\cite{kappen2016adaptive,zhang2014applications}, follow iterative procedures that progressively converge to the optimal controls.
(Note also recent neural network advances towards this direction~\cite{macris2020solving, li2020fourier}.)

\subsection{Constrained flows from time-reversed SDEs.}
Here, to circumvent the need for backward-in-time integration of the backward PDE, we express the optimal interventions $u(x,t)$ in terms of two \emph{forward} probability flows.
To that end, we consider a factorisation for the path probability density $q_t(x)$ arising from the \emph{controlled} system into two terms that account for past and future 
constraints separately~\cite{majumdar2015effective},
 \begin{equation} \label{eq:factorised1}
q_t(x) \propto \rho_t(x)  \phi_t(x). 
\end{equation}
 
In Eq.\eqref{eq:factorised1}, $\phi_t(x)$ fulfills the backward PDE (Eq.~(\ref{backward_equation})), and embodies \emph{prospective (future)} constraints to the time $t$, while $\rho_t(x)$ denotes a (non--normalised) forward probability flow that accounts for \emph{concurrent} and \emph{retrospective (past)} constraints, and is the solution of the forward PDE
 \begin{align}\label{eq:FPE2} 
\frac{\partial \rho_t(x)}{\partial t} &= {\cal{L}}_f \rho_t(x) - U(x,t) \rho_t(x)\\
=&   - \nabla \cdot\Big( f(x,t) \rho_t (x)\Big) + \frac{\sigma^2}{2} \nabla^2 \rho_t(x) - U(x,t) \rho_t(x). \nonumber
\end{align}

In the absence of path constraints $\left( U(x,t)\equiv 0 \right)$,  Eq.\eqref{eq:FPE2} reduces to the Fokker--Planck equation for the uncontrolled system. On the other hand, in the presence of path constraints  $\left( U(x,t)\neq 0 \right)$ the resulting evolution equation becomes more complicated.
By integrating Eq.\eqref{eq:FPE2} at time $t$ over a small
time interval $\delta t$, we obtain 
\begin{align}
\rho_{ t + \delta t}(x) &= e^{\delta t ({\cal{L}}_f  - U(x,t))} \rho_t(x) \label{two_stage} \\
&=  e^{-  \delta t U(x,t) } e^{\delta t {\cal{L}}_f} \rho_t(x) + O((\delta t)^2),
\end{align}
in terms of operator exponentials \footnote{In the second equality,  we considered that for small $\delta t$ the commutator of the two operators 
${\cal{L}}_f$ and $U(x,t)$ is negligible.}. This formulation admits an interpretation as the concatenation of two processes:
the propagation of the density described by the uncontrolled Fokker--Planck equation (Eq.\eqref{FPE1}), 
followed by a multiplication of the resulting density by a factor $e^{- \delta t U(x,t)}$. This second process, 
 is known in {\em filtering}  problems for stochastic dynamics~\cite{reich2013nonparametric}, where the current estimate of the system state $X_t$ results from a multiplication of the
   likelihood of the noisy observations $e^{- \delta t U(x,t)}$ with the density capturing the prior belief of the state $X_t$. Hence, we
call equation Eq.\eqref{eq:FPE2} the \emph{forward filtering equation}. In turn, the factorisation of Eq.\eqref{eq:factorised1} is reminiscent to the representation of smoothing densities for hidden Markov models as products of forward and backward messages.

The overall factorised probability flow $q_t(x)$, i.e. the probability flow characterising the evolution of the constrained system state, fulfills the Fokker--Planck equation 
 \begin{align}\label{eq:FPE_control} 
\frac{\partial q_t(x)}{\partial t} &= {\cal{L}}_g q_t(x) \\
=& - \nabla\cdot \Big( g(x,t)  q_t (x)\Big)  + \frac{\sigma^2}{2} \nabla^2 q_t(x) , \nonumber
\end{align}
 with initial condition $q_{0}(x) = \rho_0(x)$, and ${\cal{L}}_g$ denoting the Fokker--Planck operator for the optimally adjusted drift $g(x,t)$.

The factorisation of Eq.~\eqref{eq:factorised1} allows for a new representation of the optimal drift $g(x,t)$ by eliminating the backward flow $\phi_t(x)$ in favour of $\rho_t(x)$
 \begin{equation} 
g(x,t)  = f(x,t) +  \sigma^2 \left(\nabla \ln q_t(x) - \nabla \ln \rho_t(x)\right) .
\label{eq:newdrift2}
\end{equation}

 
\begin{figure*}
  \centering 
  \includegraphics[width=0.5\textwidth]{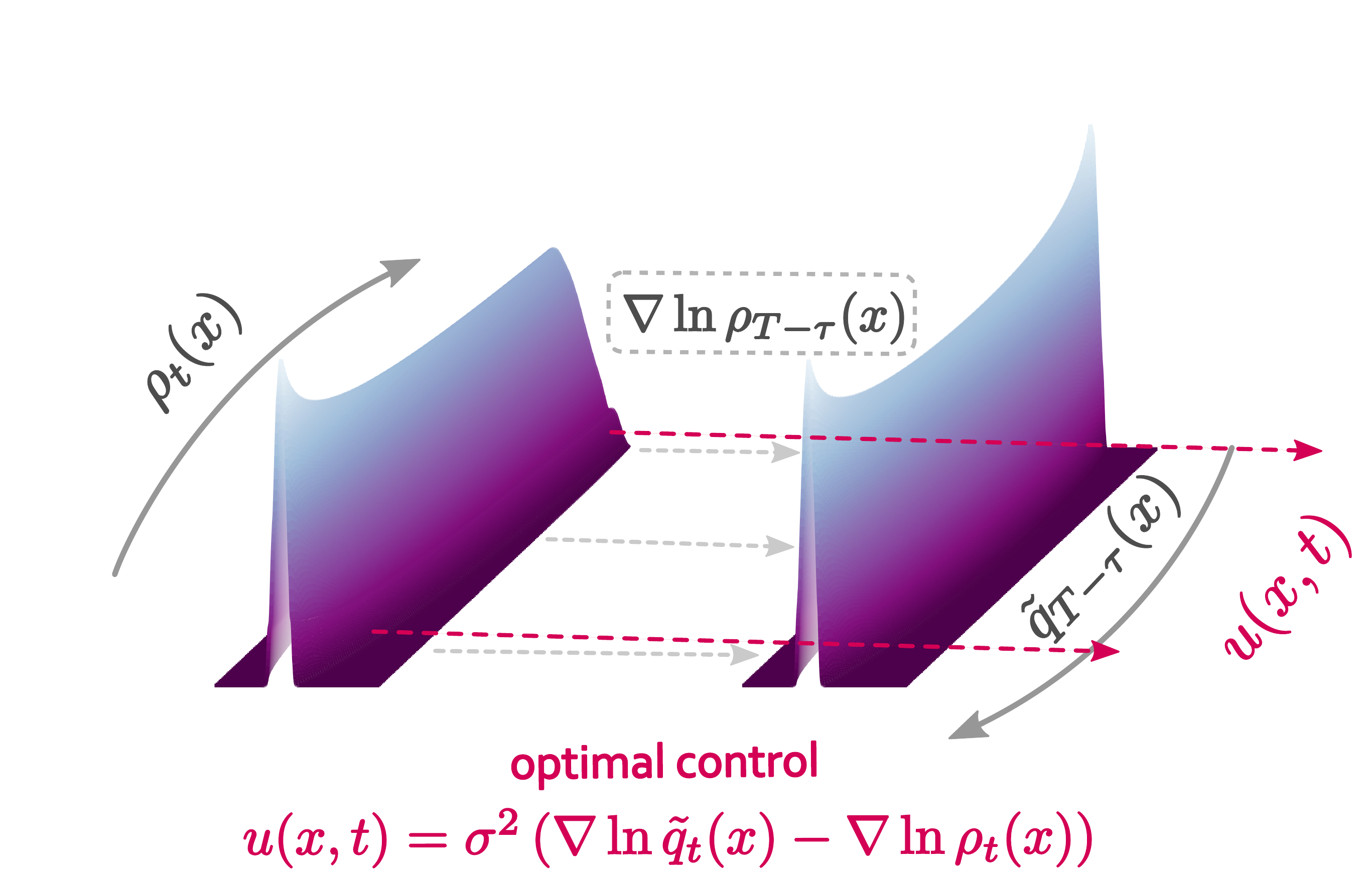}
  \caption{ \textbf{Schematic of forward and time-reversed probability flows for deriving state- and time-dependent dynamical interventions $u(x,t)$.}
  We initially sample the flow $\rho_t(x)$ for the time interval $[0,\,T]$. By employing the logarithmic gradient (\emph{score}) of $\rho_t(x)$, we evolve the time-reversed constrained probability flow $\tilde{q}_t(x)$. The optimal state- and time-dependent dynamical interventions $u(x,t)$ result from the difference of the logarithmic gradients of the two probability flows.  }   
  \label{fig:schematic}
\end{figure*}

 The new formulation of the optimal drift still requires the logarithmic gradient of the constrained flow $q_t(x)$, and therefore does not allow for direct simulation of controlled paths. 
Yet, this formulation of the optimal drift turns Eq.\eqref{eq:FPE_control} into an equation resembling a Fokker--Planck equation, but with a negative diffusion term
\begin{equation}
\frac{\partial q_t(x)}{\partial t} = \nabla\cdot \Bigg[\Big(\sigma^2\nabla \ln \rho(x, t)  - f(x,t)\Big) q_t(x)\Bigg]
-\frac{\sigma^2}{2} \nabla^2 q_t(x). 
\label{Fokker_bridge2_app}
\end{equation}
 However, by introducing the backward time variable ${\tau = T - t}$, and setting $\tilde{q}_\tau(x) = q_{T-\tau} (x)$ we obtain a 
new Fokker--Planck equation with properly signed diffusion
\begin{align}\label{Fokker_bridge3}
\frac{\partial \tilde{q}_{\tau}(x)}{\partial \tau} &= 
-\nabla\cdot \Bigg[\Big(\sigma^2\nabla \ln  \rho_{T-\tau} (x)  - f(x, T-\tau)\Big)  \tilde{q}_{\tau} (x)\Bigg] \\ \nonumber
&+ \,\, \frac{\sigma^2}{2} \nabla^2 \tilde{q}_{\tau} (x) , 
\end{align}
with initial condition $\tilde{q}_{0} (x) \propto \rho_T(x) \chi(x)$.

 Hence, we have represented the optimal control 
\begin{equation} 
u^*(x,t) = \sigma^2 \Big(\nabla \ln \tilde{q}_{T- t} (x) - \nabla \ln \rho_t(x)\Big) ,
\label{eq:backward_drift2}
\end{equation}
as the difference of the logarithmic gradients of two forward probability flows (\emph{ score--functions }).

\if False
To obtain optimal dynamical interventions $u^*(x,t)$, we sample the two probability flows $\rho_t(x)$ and $\tilde{q}_t(x)$, and obtain the optimal dynamic adjustments $u^*(x,t)$ by computing the difference of the logarithmic gradient of the two flows.
By incorporating the time-dependent dynamical perturbation $u^*(x,t)$ into the dynamics of the considered system, we successfully drive the controlled trajectories to the predefined targets respecting thereby the imposed path constraints.
\fi

 Hence, we have represented the optimal control 
\begin{equation} 
u^*(x,t) = \sigma^2 \Big(\nabla \ln \tilde{q}_{T- t} (x) - \nabla \ln \rho_t(x)\Big) ,
\label{eq:backward_drift2}
\end{equation}
as the difference of the logarithmic gradients of two forward probability flows ({\em score--functions}).
This result suggests
a possible numerical strategy for obtaining the optimal interventions:  
First, solve the forward filtering equation
for $\rho_t(x)$  (Eq.~\eqref{eq:FPE2}) using e.g. a particle filtering approach. Then approximate the logarithmic gradients $\nabla \ln  \rho_{T-\tau} (x)$
based on a sufficiently large number of stochastic particle paths. After this, 
simulate a number of random trajectories of the SDE which corresponds to the Fokker--Planck equation (Eq.~\eqref{Fokker_bridge3})
and use the trajectories to approximate the score $\nabla \ln \tilde{q}_{T- t} (x)$.
In fact, a similar approach was recently used to solve so--called {\em Schr\"odinger bridge} problems~\cite{de2021diffusion}. The latter can be 
also be understood as specific control problem for SDEs with only control energy costs (i.e. $U(x)\equiv 0$ and $\chi(x) =1$) in the cost functional (Eq.\eqref{eq:cost_to_go}), but in addition the
probability densities $q_0(x)$ and $q_T(x)$ at initial and final times are specified as extra constraints.
In contrast to these approaches, we apply and generalise a recent {\em deterministic}, particle framework for solving Fokker--Planck 
equations introduced by the authors in \cite{Maoutsa_2020}
to solve generic path integral control problems of the type of Eq.~\eqref{eq:cost_to_go}. This new technique avoids stochastic path sampling
and reduces significantly temporal fluctuations, delivering thereby accurate Fokker--Planck equation solutions for relatively low number of employed particles~\cite{Maoutsa_2020}. 
In addition, the computation of the logarithmic gradients is already an integral part of the method for computing the deterministic particle dynamics.

\subsection{Deterministic particle flow (DPF) control}\label{sec:dfp}
To sample the forward densities $\rho_t(x)$ and $\tilde{q}_t(x)$, we build on the idea that a Fokker--Planck equation can be rewritten as a \emph{Liouville equation}~\cite{gardiner2009stochastic} 
for an ensemble of deterministic dynamical systems where the logarithmic gradient of the ensemble density acts as an additional force.
In particular, for an SDE with drift $f(x,t)$ and diffusion $\sigma$, we rewrite the Fokker--Planck equation (Eq.\eqref{eq:free_SDE}) for
the probability density $p_t(x)$ of the system state in the form (see~\cite{Maoutsa_2020} for details)  
\begin{equation}
\label{eq:flow}
\frac{\partial p_t(x)}{\partial t} = - \nabla \cdot \left[  \left(  f(x,t) - \frac{\sigma^2}{2}  \nabla \ln p_t(x)\right)   p_t(x) \right].
\end{equation}
For a known density $p_t(x)$, this equation describes the evolution of an ensemble of independent systems with 
each ensemble member following the deterministic dynamics:
\begin{equation}
\frac{d {X}_t}{dt} = f({X}_t,t) - \frac{\sigma^2}{2}  {\nabla} \ln p_t({X}_t). 
\end{equation}
Note that here individual trajectories $X_{0:T}$ are distinct from solutions of the underlying SDE, since each ensemble individual follows pure deterministic dynamics.

To obtain a solution of the Fokker--Planck equation, we approximate the density $p_t(x)$ by an empirical distribution of $N$ ensemble members (\emph{'particles'}) $\{X_t^{(i)}\}_{i=1}^N$ via
\begin{equation}
\hat{p}_t(x) \approx \frac{1}{N}\sum_{i=1}^N \delta \left(x - X_t^{(i)}\right).
\label{particle_density}
\end{equation}
Based on this empirical representation of the density $p_t(x)$, we approximate its logarithmic gradient with a statistical {\em estimator} $\hat{\nabla} \ln \hat{p}_t(X^{(i)}_t)$,
obtained from the solution variational formulation of the score function (see Supplementary Information). Thus, we express the resulting dynamics of individual
particles in terms of a system of ordinary differential equations (ODEs)
\begin{equation} \label{eq:deter_dynamics}
    \frac{d X^{(i)}_t}{dt} = f(X^{(i)}_t,t) - \frac{\sigma^2}{2}  \hat{\nabla} \ln \hat{p}_t(X^{(i)}_t). 
\end{equation}

While this approach is sufficient to solve control problems without path costs ($U(x,t) \equiv 0$), the 
extra sink term $-U(x,t) \rho_t(x)$ in the forward filtering PDE (Eq.\eqref{eq:FPE2}) in the presence of path constraints requires an additional numerical technique.
Thus, to incorporate path costs, we employ the formulation of the two stage process given by Eq.\eqref{two_stage}, and combine our Fokker-Planck deterministic particle solver with a deterministic particle filter method, the
{\em ensemble transform particle filter}~\cite{reich2013nonparametric}.
To simulate such a two stage process for each small time interval $\delta t$,  we first 
propagate the particles following the dynamics of Eq.\eqref{eq:deter_dynamics} to auxiliary positions $Y^{(i)}_t$ and 
assign to each particle $i$ a {\em weight} $\Omega_i$
\begin{equation}
\Omega_i(t) \propto e^{-\delta t U(Y^{(i)}_t,t)}.
\label{reweight}
\end{equation}
To transform the weighted particles to unweighted ones, we employ the {\em ensemble transform particle filter}~\cite{reich2013nonparametric}. This method solves an Optimal Transport~\cite{villani2009optimal} problem to provide the minimal necessary {\em deterministic} shifts required to transform an ensemble of weighted particles into an ensemble of uniformly weighted ones, by maximising the covariance between the two ensembles (see Supplementary Information).


\subsection{Guiding probability flows to extreme terminal states} \label{sec:extreme_points}

In settings where the terminal target state lies outside of the typical values of the uncontrolled system, the sampled forward flow $\rho_t(x)$ fails to provide sufficient evidence in the vicinity of the terminal point. Thereby the ensuing logarithmic gradient estimation $\nabla \log \rho_t(x)$ is inaccurate.

For conservative systems and for terminal constraints defined by a delta function, i.e. $\chi(x) = \delta(x - x^*)$, we address this issue by proposing an additional modified forward sampling that incorporates the extreme terminal constraint in the forward dynamics. To that end, we employ a \emph{$d$-dimensional Brownian Bridge (BB)} dynamics. Brownian bridges are essentially Brownian motions, i.e. diffusions with vanishing drift $f(x) \equiv 0$, with a fixed terminal state $x^*$ (see Supplementary Information). 

To maintain the correct path statistics, we employ the Girsanov's change of measure formula to reweight the biased forward paths.
More precisely, we obtain the correct path probability measure of the controlled process $\mathbb{P}^{BB}_f$ by reweighting the Brownian bridge path measure 
$\mathbb{P}^{BB}_0$ with the likelihood (Supplementary Information)
\begin{equation}
\frac{d\mathbb{P}^{BB}_f}{d\mathbb{P}^{BB}_0} (X_{0:T}) 
\propto \exp\left[ - \int_0^T U_{BB}(X_t) dt \right],
\end{equation}
with
\begin{equation}
U_{BB}(x) = \frac{1}{2\sigma^2}\left(f^2(x) + \sigma^2 \nabla\cdot f(x)\right).
\label{U_bridge}
\end{equation}

 Hence, to simulate constrained paths of an SDE with drift $f(x)$ and extreme terminal constraints, we transform the extreme terminal constraint to a path constraint $U_{BB}(x)$ for an appropriate Brownian bridge process that already incorporates the terminal state $x^*$. In particular, we employ the modified forward equation 
 \begin{equation}
\frac{\partial \rho_t(x)}{\partial t} = {\cal{L}}_{f_0} \rho_t(x) - U_{BB}(x) \rho_t(x) ,
\label{forbridge}
\end{equation}
that generates paths with correct statistics that reach the terminal target by imposing the path constraint $U_{BB}(x)$ to the Brownian bridge forward dynamics  with drift $f_0$. We term this variant of our framework \emph{guided Deterministic Particle Flow} (\gDPF) control.



\section*{Evaluation of optimal dynamical interventions}
To illustrate our formalism in action, we computed optimal intervention protocols for biological systems by employing the proposed deterministic particle framework, and compared the obtained controls to those computed with the \emph{Path integral cross entropy method} (\PICE) (see Supplementary Information and ~\cite{kappen2016adaptive}). We tested our method on systems of increasing complexity and dimensionality, with conservative and non-conservative forces, as well as in settings with terminal, path, or collective state constraints. 

To design the optimal interventions $u(x,t)$ we employed the presented method in two alternative variants: $i)$ the \emph{deterministic particle flow control} (\textbf{DPF}), where the forward density follows the dynamics of Eq.~\eqref{eq:FPE2}, and $ii)$ the \emph{guided deterministic flow control} (\textbf{gDPF}), in which the forward density evolves according to an appropriately reweighted Brownian bridge dynamics, as described in Section~\ref{sec:extreme_points}.

 To evaluate the quality of the obtained controls, we considered the design of optimal interventions for artificially manipulating molecular phenotypes on adaptive landscapes (Section~\ref{res:evol}), for inducing state transitions to multistable conservative and non-conservative systems (Section~\ref{res:cell}), and for synchronising finite-size networks of Kuramoto phase oscillators (Section~\ref{res:kura}). We quantified the quality of the identified interventions in terms of employed control energy ($\|u(x,t)\|^2_2$), reflecting the optimality of the computed control, as well as in terms of deviations from terminal ($(x^* - X_T)^2$) and path constraints ($U(X_t,t)$), characterising thereby the effectiveness of the devised interventions to enforce the intended constraints. Unless explicitly mentioned otherwise, all metrics were evaluated by considering $1000$ independent stochastic trajectories controlled with each method.

\subsection{Controlling state transitions of conservative and non-conservative systems.}\label{res:cell}
\begin{figure*}[ht!]
  \begin{overpic}[width=\textwidth]{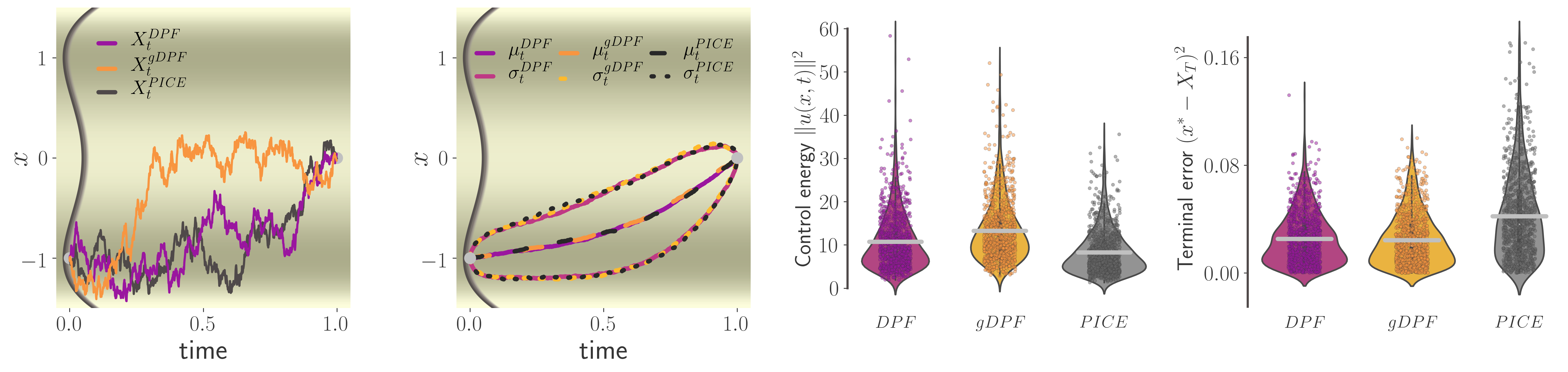}
\put(-1,21){\color{black}{\Large\textbf{ a.}}}
\put(24,21){\color{black}{\Large\textbf{ b.}}}
\put(48,21){\color{black}{\Large\textbf{ c.}}}
\put(71,21){\color{black}{\Large\textbf{ d.}}}
\end{overpic}
\end{figure*}

\begin{figure*}[ht!]
   \begin{overpic}[width=\textwidth]{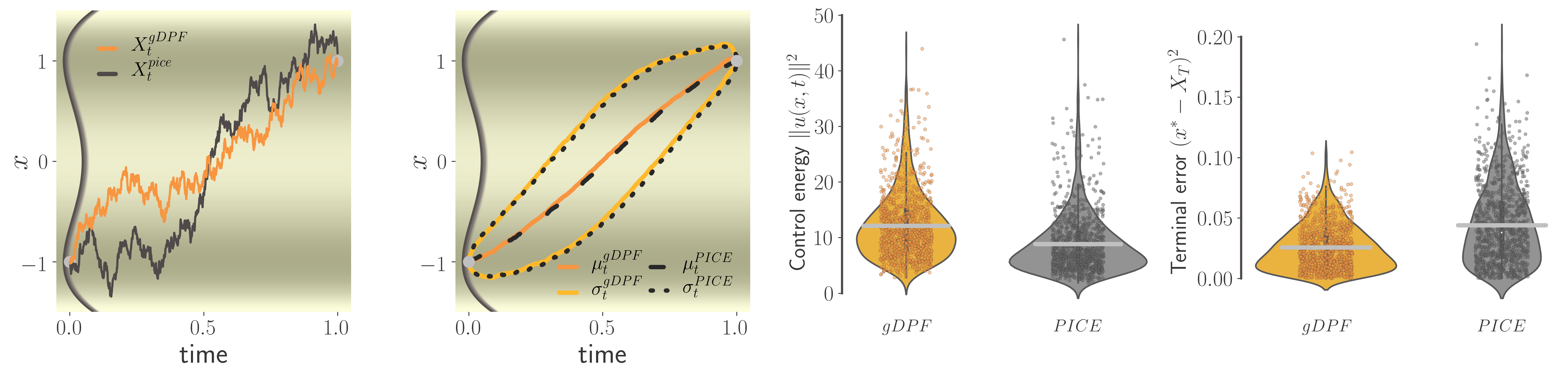}
\put(-1,21){\color{black}{\Large\textbf{ e.}}}
\put(24,21){\color{black}{\Large\textbf{ f.}}}
\put(48,21){\color{black}{\Large\textbf{ g.}}}
\put(71,21){\color{black}{\Large\textbf{ h.}}}
\end{overpic}
  \caption{ \textbf{Equivalence of dynamical interventions delivered by deterministic particle flow control (\emph{DPF}), guided deterministic particle flow control (\emph{gDPF}), and PICE, for typical (upper) and extreme (lower) terminal conditions.} \textbf{(upper) (a.)} Controlled trajectories simulated by employing dynamical interventions delivered by deterministic particle flow control  (\emph{magenta}), guided deterministic particle flow control  (\emph{orange}), and path integral cross entropy method (\emph{grey}). All trajectories started from initial point $x_0=-1$ (left silver circle) and reached the target $x^*=0$ at time $T=1$. \textbf{(b.)} Transient mean $\mu_t$ and standard deviation $\sigma_t$ over $1000$ independent trajectories controlled with each framework.\textbf{(c.)} Total control energy $\| u(x,t) \|^2$, and \textbf{(d.)} deviation from terminal point for $1000$ independent controlled trajectories with interventions computed according to each framework. Grey horizontal lines in (c.) and (d.) denote mean values over all realisations. The proposed methods result in slightly more expensive control costs, but are more consistent in precisely reaching the terminal state.   \textbf{(lower) } Same as upper row for target $x^*=1$  only for \gDPF ~and \PICE. Here \DPF ~is not applicable since the forward probability flow does not reach the extreme target point $x^*=1$.   }   \label{fig:double_well}
\end{figure*}
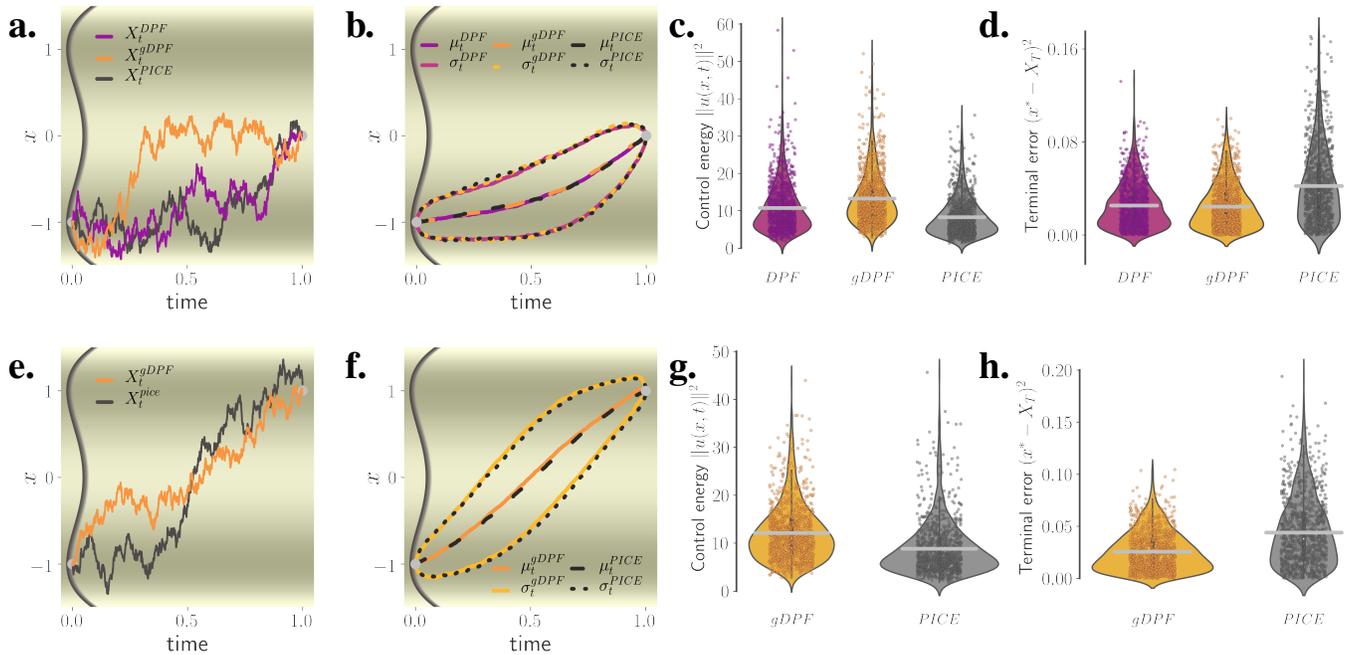

 We employed the proposed framework to devise interventions that reliably induced switching between stable states in a time constrained scheme for an one-dimensional conservative system and for a two-dimensional non-conservative one. For both settings, the unconstrained system either performs the state transition in a time unreliable way, or completely fails to reach the target, when the transition paths to that state strongly deviate from typical trajectories (Figure~\ref{fig:double_well}).

For the one-dimensional bistable system ($f(x)= 4\, x - x^3 $) starting from the stable state at $x_0=-1$, 
we provided optimal interventions under the objective of driving the system towards a predefined target $x^*$ at time $T=1$. We applied both variants of the presented method \big(\DPF ~(Section~\ref{sec:dfp}) and \gDPF ~(Section~\ref{sec:extreme_points})\big), and explored two complementary scenarios: one with typical, $x^*=0$ (Figure~\ref{fig:double_well}a.-d.), and one with extreme, $x^*=1$ (Figure~\ref{fig:double_well}e.-h.), target states for the uncontrolled system at time $T$. Notice that the state $x^*=0$  is unstable, but outside of the basin of attraction of the initial state $x_0$.

\textbf{Identified interventions successfully drive the system to target states.}
For the typical target state $x^*=0$, all three employed methods (\textbf{DPF:~\emph{magenta}}, \textbf{gDPF:~\emph{yellow}}, \textbf{PICE:~\emph{grey}}) successfully biased the controlled system towards the target $x^*$ (Figure~\ref{fig:double_well}a.). In fact, the distributions of simulated independent controlled trajectories delivered by each framework completely agreed throughout the entire time interval $[0,\,T]$ (Figure~\ref{fig:double_well}b.).

Considering the control energy dissipated by each method, \DPF, on average, provided slightly larger interventions (Figure~\ref{fig:double_well}c.). 
Yet, both variants of our approach (\DPF ~and \gDPF) induced control trajectories that were consistently more \emph{exact} in reaching the terminal state (Figure~\ref{fig:double_well}c.). 

Comparing the distributions of terminal errors, \DPF ~was both more accurate in reaching the target, as mediated by a smaller average value over the $1000$ realisations (grey bar), but also more precise, as indicated by the smaller dispersion of terminal errors around the average. This demonstrates that although \DPF ~slightly overestimated the required controls, it provided sufficient force to lead the trajectories faithfully onto the target. In contrast, \PICE ~relatively underestimated the necessary interventions, resulting in more energy efficient controls that nevertheless moderately deviated from the target.

Importantly, these results suggest that a \emph{single} iteration of either variants of the proposed method (\DPF ~or \gDPF) provide comparable controls to the \emph{iterative} \PICE ~framework (grey).

For the extreme terminal state $x^*=1$, the guided probability flow deterministic control (\gDPF) and the path integral cross entropy method (\PICE) successfully pushed the system to the target (Figure~\ref{fig:double_well}e.)). The distributions of controlled trajectories (Figure~\ref{fig:double_well}f.) from the two frameworks showed complete agreement, while the control costs and terminal state precision were qualitatively similar to those obtained for the typical target state.

Notice that the deterministic particle flow control (\DPF), the simple variant of our method, is inappropriate for this setting for a reasonable number of employed particles representing the forward flow. Since the target is an extreme system state for time $T$, it is highly unlikely that particles representing the forward flow $\rho_t(x)$ will reach the target $x^*$. Thus the estimation of logarithmic gradients of the forward flow $\rho_t(x)$ in the vicinity of the target will be inaccurate, since the particles will not provide sufficient evidence for gradient estimation in that region.

Departing from gradient systems, onward we consider triggering transitions between stable states in a time reliable way for a non-conservative system.
To that end, we employed \DPF ~for controlling a two dimensional phenomenological model of the function of a cell fate division module of a gene regulatory network~\cite{huang2007bifurcation}
with self-excitation and cross inhibition (see Supplementary Information). We 
 applied our method to induce transitions to the system between its co-existing stable states.
 
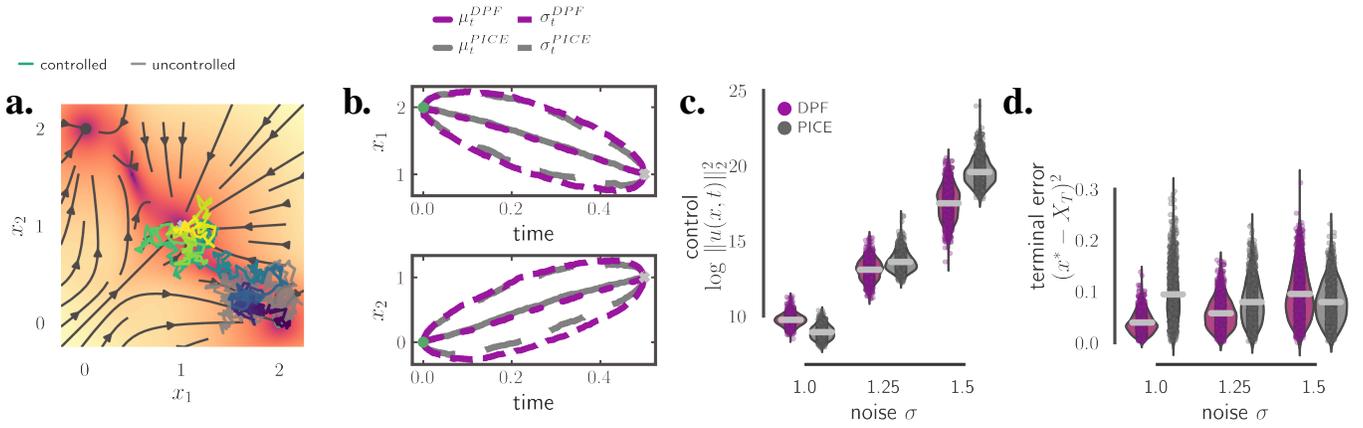
\begin{figure*}[ht!]
  \begin{overpic}[width=\textwidth]{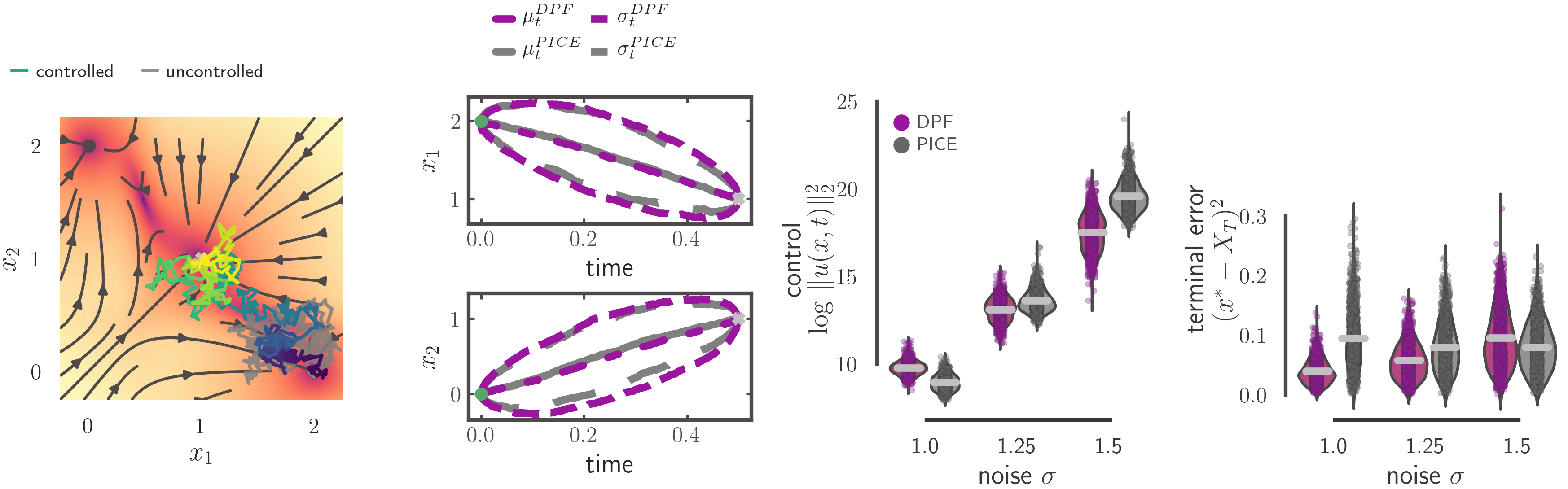}
\put(-1,23){\color{black}{\Large\textbf{ a.}}}
\put(24,23){\color{black}{\Large\textbf{ b.}}}
\put(49,23){\color{black}{\Large\textbf{ c.}}}
\put(73,23){\color{black}{\Large\textbf{ d.}}}
\end{overpic}
  \caption{ \textbf{Optimal interventions effectively drive the non-conservative system to the target state for different noise amplitudes.} \textbf{ (a.)} Individual trajectory controlled by \DPF ~(green-yellow) starting from stable state $\textbf{x}_0=(1.996,\, 0.4)$ successfully reaches the target $\textbf{x}^*= (1,1)$ at $T=0.5$, while the uncontrolled trajectory (grey) fails to leave the basin of attraction of $\textbf{x}_0$. \textbf{(b.)} Agreement between controlled densities of $1000$ independent controlled trajectories driven by \DPF ~(magenta) and \PICE ~(grey) for noise amplitude $\sigma=1.5$. \textbf{(c.)} Dissipated control energy, and \textbf{(d.)} terminal error for increasing noise strength $\sigma$ for the two control frameworks. For increasing noise \DPF ~delivers more efficient, but moderately less precise controls than \PICE.\footnote{Further parameters: particle number: $N=600$ inducing point number: $M=20$.}  }  \label{fig:c_eleg} 
\end{figure*}

Similar to the conservative setting examined previously, the \DPF ~successfully mediated the necessary interventions for the transition between the stable states (Figure~\ref{fig:c_eleg} a.). We considered three noise conditions with $\sigma=\{1., 1.25, 1.5\}$ and compared again the deterministic particle flow control (\DPF) with the path integral cross entropy method (\PICE). The transient statistics computed over $1000$ controlled trajectories for each framework agreed for all noise conditions (Figure~\ref{fig:c_eleg}b.) and Supplementary Information Figure). 

Control costs and terminal error precision were comparable for the two approaches, with \DPF ~performing slightly better in terms of dissipated control energy for increasing noise strength (Figure~\ref{fig:c_eleg}c.). Both methods had comparable control accuracy in reaching the target that deteriorated moderately for increasing noise amplitude. However, while \DPF ~was considerable more accurate and precise for the low noise conditions, for larger noise amplitudes the accuracy of both methods in reaching the terminal point $x^*$ became comparable (Figure~\ref{fig:c_eleg}d.).

Taken together, \DPF ~(and \gDPF ~where necessary) successfully provided the necessary controls for reaching the targets in both conservative and non-conservative systems and under various noise conditions. The provided interventions where on par with the established iterative \PICE ~framework in terms of dissipated control energy, and moderately more precise in reaching the target.

\subsection{Evolutionary control through artificial selection.}\label{res:evol}

\begin{figure}[ht!]
  \centering 
  \includegraphics[width=0.45\textwidth]{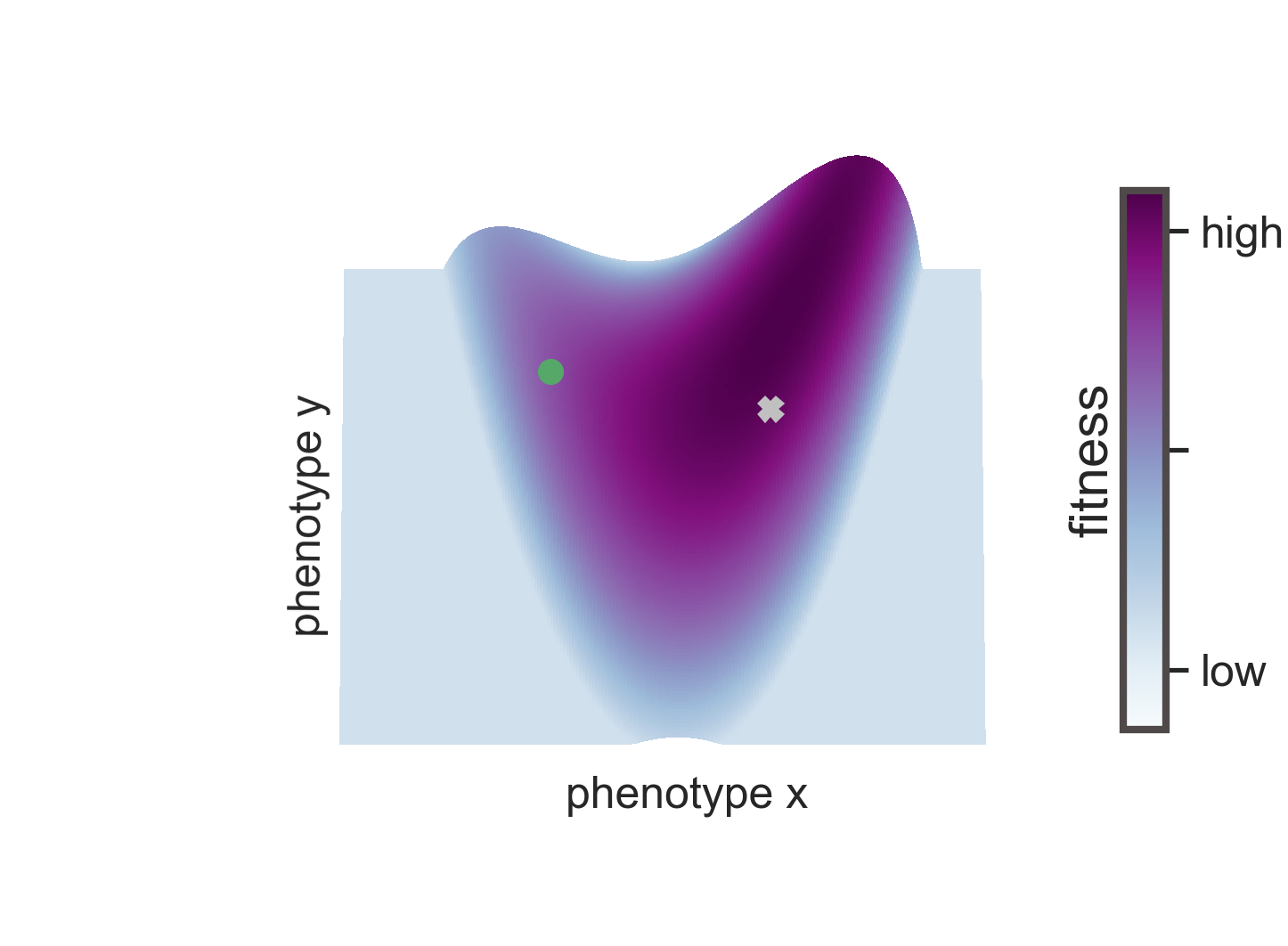}
  \caption{ \textbf{The phenotypic landscape .}      }  \label{fig:adaptation_landscape} 
\end{figure}

\begin{figure*}[ht!]
  \centering 
  \includegraphics[width=0.951\textwidth]{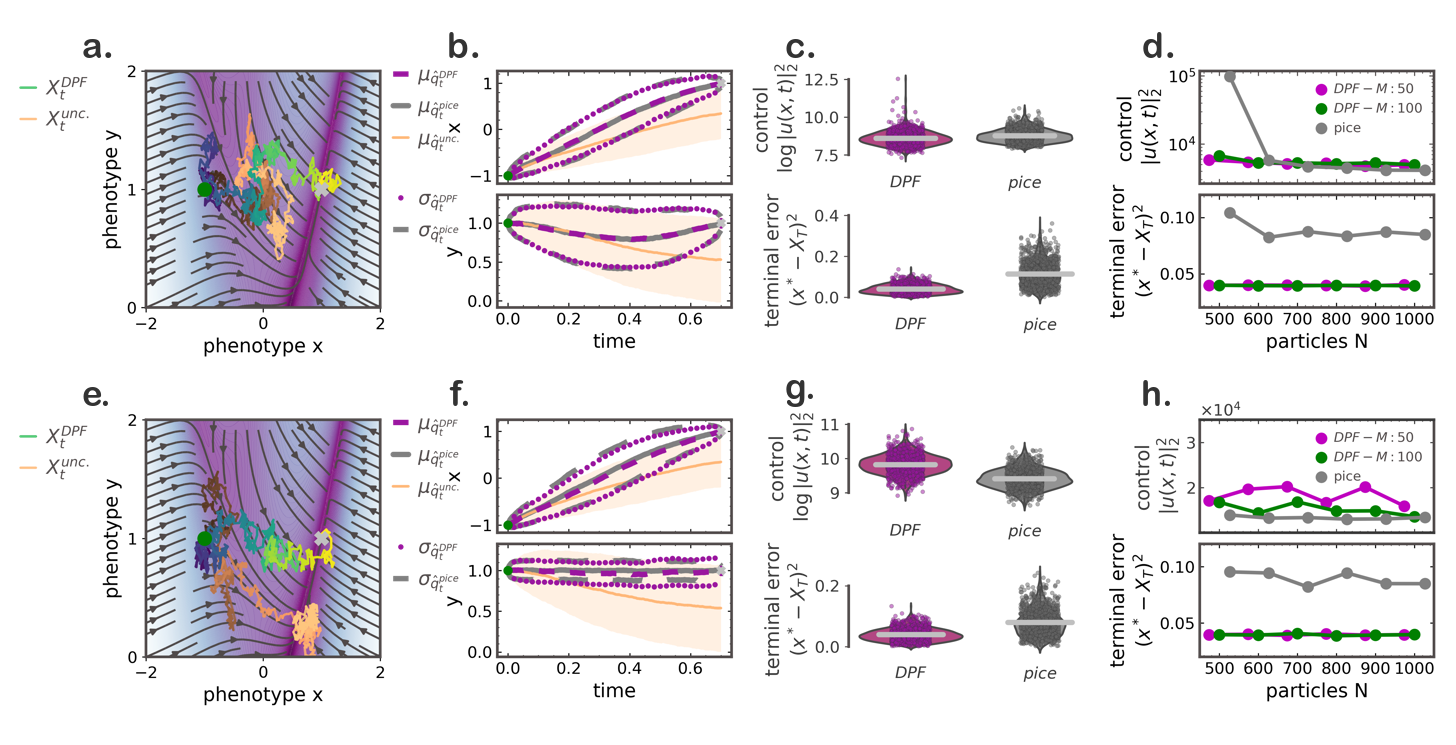}
  \caption{ \textbf{Deterministic particle flow (DPF) control provides optimal interventions to drive evolution to target state (\emph{grey cross}).} \textbf{(a.)} A controlled trajectory starting from phenotypic state $(-1,1)$ reaches the target state $(1,1)$ at time $T=0.7$ (\emph{blue-yellow line}), while an uncontrolled trajectory remains in the vicinity of initial state during the same time interval (\emph{orange line}). \textbf{(b.)} Mean and standard deviation of the marginal densities of $1000$ controlled trajectories employing interventions computed with our framework DPF (purple) and with PICE (grey). Orange line indicates mean of $1000$ uncontrolled trajectories, while shaded area captures the associated standard deviation. For estimating the controls we employed $N=400$ particles for DPF, and $N_{pice}=500$ for PICE. \textbf{(c.)} Comparison of \textbf{(upper)} (logarithmic) control energy $\|u(x,t)\|^2_2$, and \textbf{(lower)} deviation from the terminal point $\|X_T - x^*\|^2$  for each controlled trajectory with interventions computed according to DPF (magenta) and PICE (grey). \textbf{(d.)} (Logarithmic) control energy (upper) improves moderately, and terminal error (lower) remains constant for increasing particle number $N$. The number of inducing points in the logarithmic gradient estimation conveys negligible difference in control energy and terminal error (inducing point number magenta: $M=50$, green: $M=100$ ). Grey line indicates the performance of PICE in the same setting. \textbf{(e.-h.)} Same as (a.-d.) with additional path constraint $U((x,y),t)= (y-1)^2$.
  }   \label{fig:nonpath_main}
\end{figure*}


 Building upon the evolutionary stochastic control formalism recently introduced by Nourmohammad et. al \cite{nourmohammad2021optimal}, we employed deterministic particle flow control to devise artificial selection protocols for molecular phenotypes that optimally drive evolution to desired phenotypic states. In this setting experimentally imposed path constraints become relevant for preventing undesirable outcomes on covarying phenotypes.\\

\if False
Phenotypic selection by natural adaptive forces of an evolving population can be thought of as the mean fitness among all individuals in the population evolving according to a landscape gradient until it reaches equilibrium, i.e. the peak of the landscape.
The fitness of each phenotypic frequencies in the population is denoted by the height of the landscape. The distribution of phenotypes in the population is denoted by a point on the landscape and according to Wright natural selection forces push the population towards the nearest topographic local maximum. Thus the direction and rate of evolution depend on the the geometry of the phenotypic landscape. local maxima are equivalent to to ecological niches for a population.
\fi

For an evolving population, the main evolutionary drivers comprise fitness and mutation forces that continuously adjust the composition of phenotypes in the population, while genetic drift perturbs the whole process stochastically. 
We described the evolution of the \emph{mean phenotypes} $\mathbf{dx}$ of the population by the overdamped Langevin equation~\cite{lande1976natural}
\begin{equation} \label{eq:pheno1}
    d\mathbf{x} = C \cdot\nabla F(x) dt + \sigma d\mathbf{W},
\end{equation}
with $F(x)$ denoting the adaptive fitness landscape in the presence of natural selection~\cite{fisher1958genetical}, 
and $\sigma$ the noise amplitude that rescales the genetic drift, i.e. the stochastic term, according to the population size $n$ and the covariance matrix $C$, $\sigma = C^{1/2} n^{-1}$ . The gradient of the landscape $ f(x) = C \cdot \nabla F(x)$ captures the adaptive pressures under natural selection. 

In contrast to commonly employed assumptions of spherically symmetric adaptive landscapes, to demonstrate the richness of our method and supported by empirical findings~\cite{kingsolver2001strength}, we considered an asymmetric rugged landscape~\cite{barton2007evolution}. Such landscapes may arise in small sized populations with small variance and multi-modal fitness functions for each individual~\cite{whitlock1995variance}. 
For simplicity we consider the covariance matrix $C$ constant, given the much smaller timescales upon which its fluctuations unfold~\cite{nourmohammad2013evolution}, and its weak dependency on the evolutionary selection strength~\cite{held2014adaptive}.

The dynamics of Eq.(\ref{eq:pheno1}) describe the evolution of populations in the presence of natural selection towards an evolutionary optimum, captured by the maximum of the adaptive landscape, adhering thereby to physiological and environmental constraints. 

To study the outcomes and dynamics of adaptive evolution, intervention protocols are required that drive phenotypes towards non-evolutionary optimum states $\mathbf{x^*}$, or through evolutionary trajectories that deviate from landscape gradients. These interventions are implemented through artificial selection, which, following~\cite{nourmohammad2021optimal}, we formulate as a time- and state- dependent perturbation $\mathbf{u}(x,t)$ to the natural selection, and apply \DPF ~to obtain the necessary controls.



\textbf{Single-objective and multi-objective directed evolution.} For a system with multiple covarying phenotypes evolving on the landscape $F(x,y) = ((1-x)^2 + (y - x^2)^2)$ (Figure~\ref{fig:adaptation_landscape}), changes along one phenotypic axis are often accompanied by changes along covarying phenotypic traits as dictated by the landscape gradient. Thus attempts to bias and enhance selective forces towards a specific phenotype, may lead to undesired variations along the covarying traits. To demonstrate this, we controlled phenotypic trajectories initiated at state $\mathbf{x}_0= (-1,1)$ with target state $\mathbf{x}^*=(1,1)$. The designed interventions by \DPF ~with only terminal constraints $\chi(\mathbf{x})= \delta(\mathbf{x} - \mathbf{x}^*)$
successfully drove the system to the intended target (Figure~\ref{fig:nonpath_main}a.-b.). Nevertheless, although the initial and terminal state along the second dimension remained the same, as expected, without introducing additional path constraints, the phenotypic trait along the second dimension ($y$) underwent considerable transient fluctuations as indicated by the non-constant mean of the simulated paths ($\mu_{\hat{q}_t}$), as well as by the increasing dispersion of paths from the mean, as captured by the standard deviation $\sigma_{\hat{q}_t}$ (Figure~\ref{fig:nonpath_main}b.). 

To limit the fluctuations along the covarying second phenotypic trait, we introduced path constraints \[U((x,y),t) = 10^3\,(y-1)^2\] that penalised transient deviations from the intended value of the second trait $y$ (Figure~\ref{fig:nonpath_main}e.). The necessary interventions, identified by \DPF ~with path constraints, successfully controlled the system towards the predefined target, reducing thereby considerably the fluctuations along the second axis (Figure~\ref{fig:nonpath_main}e. and f.). 
Compared to the path integral control framework (\PICE), for both scenarios, our method delivered results with comparable dissipated control energy and considerably more precise in terms of terminal errors (Figure~\ref{fig:nonpath_main}e. and f.).

Evaluating the performance of both methods for increasing particle number $N$ employed for the estimation of the required interventions $u(x,t)$, revealed that \DPF ~provided controls on par with \PICE ~already for $N=500$ particles and, by design, only with a \emph{single} iteration.
More precisely, the proposed method presented a relatively stable performance for increasing particle number, with small improvement in terms of exerted control energy, whereas the path integral cross entropy method improved substantially when more particles where employed in the computations, and consequently matched in performance our approach. Considering the deviation from the terminal state, \DPF ~was consistently more precise and accurate in reaching the target as indicated both by considerably smaller terminal errors (Figure~\ref{fig:nonpath_main}d. and h.), and by smaller deviations of individual terminal states around the average (Supplementary Figure~\ref{fig:nonpath_main}). Both the exerted control energy and terminal error do not show considerable improvement for increasing inducing point number employed in the logarithmic gradient estimator (magenta for $M=50$; green for $M=100$).

\if False

\textbf{Employing path constraint limits the fluctuations along the direction of the constrained phenotype for both frameworks}

Starting from an initial phenotypic state $x_0 =-1$, we demonstrate the performance of the two variants of our method against PICE when the target is a \emph{typical} $x^* = 0$ and an \emph{extreme} state $x^* = 1$.

We evaluate the identified time- and state- dependent optimal interventions of each algorithm by simulating $1000$ independent controlled trajectories for each method.  

State interventions delivered by the deterministic particle flow control (\DPF) and the guided deterministic particle flow control (\gDPF) both successfully drive the phenotypic state to the predefined target state $x^*$ at time $T$, when $x^*$ is a typical state for the considered system as shown by individual realisations of controlled trajectories (magenta (\DPF) and yellow paths(\gDPF)) (Fig.\ref{fig:double_well} (a.) ). All three frameworks resulted in identical statistics when considering the entire ensemble of controlled paths with their transient mean ($\mu_X$) and transient ($\sigma_X$) of the resulting path probability densities agreeing throughout the entire interval $[0,\,T]$. 

To evaluate the quality of interventions computed by the three frameworks we compared the total control energy $\int_0^T\|u(x,t) \|^2 dt$ spent (Fig.\ref{fig:double_well} (c.) ), capturing the optimality of the estimated interventions, and the deviation from the terminal point $\|X_T - x^*\|^2$ for each controlled realisation (Fig.\ref{fig:double_well} (c.) ), indicating the accuracy of the delivered controls.
While on average the

The statistics (mean and standard deviation) of the distribution of $1000$ controlled trajectories with our framework, indicated as $\hat{q}_t(x)$, are in complete agreement with the statistics of controlled trajectories , and with the reversed-time sampled flow $q_t(x)$ and the

\fi

\subsection{Controlling collective states: synchronisation control of stochastic Kuramoto oscillators} \label{res:kura}

\begin{figure*}[ht!]
  \begin{overpic}[width=\textwidth]{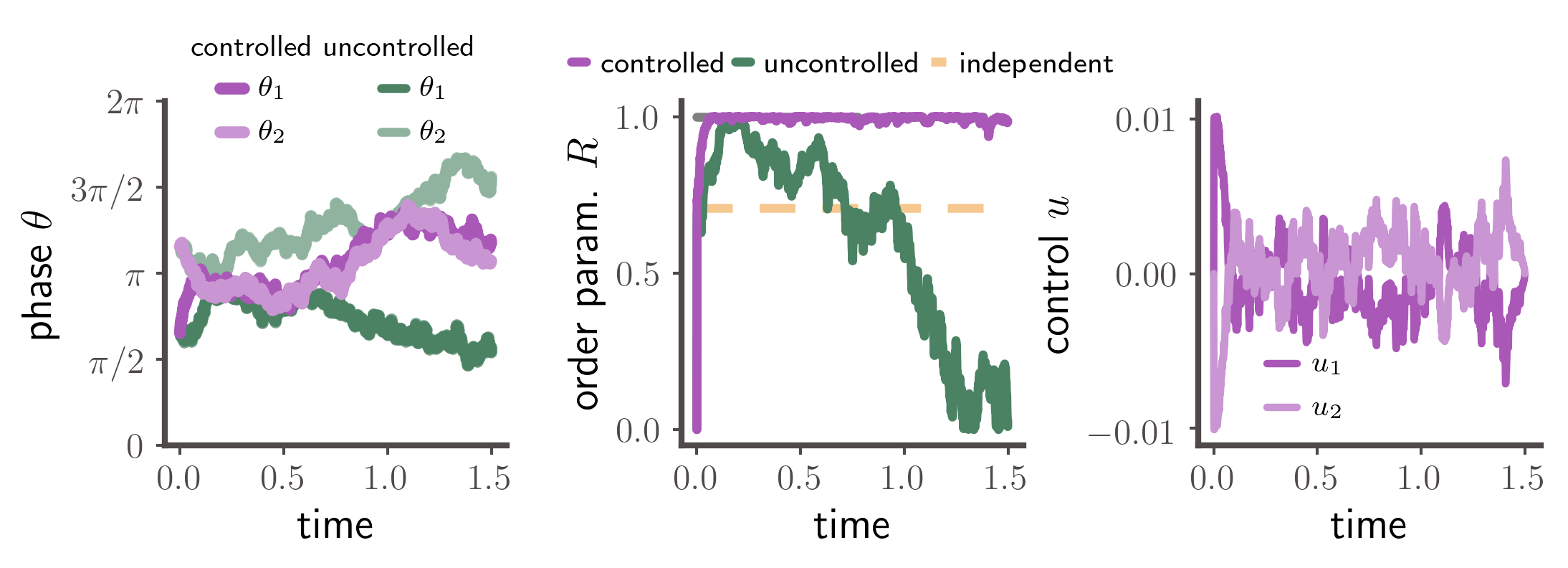}
\put(2,28){\color{black}{\Large\textbf{ a.}}}
\put(34,28){\color{black}{\Large\textbf{ b.}}}
\put(66,28){\color{black}{\Large\textbf{ c.}}}
\end{overpic}
  \caption[two_kuramoto]{ \textbf{Synchronisation control of two coupled Kuramoto phase oscillators.}  \textbf{(a.)} Evolution of phases ($\theta_i$) of two controlled (\emph{purple}) and two uncontrolled (\emph{green}) Kuramoto oscillators mutually coupled with coupling $J=1.2$ and noise amplitude $\sigma=1.0$.
   \textbf{(b.)} Evolution of Kuramoto phase-coherence order parameter $R$ for the controlled (\emph{purple}) oscillators indicates a fast transition to synchrony (R=1), while, the identical uncontrolled oscillators become progressively incoherent, indicated by a strongly fluctuating order parameter (\emph{green}). The orange line denotes the expected long time average value of the order parameter for non-interacting oscillators considering finite size scaling effects. The grey line marks the level of $R=1$ indicating a completely synchronous state. \textbf{(c.)} Control input provided to each oscillator.  }\label{fig:kuramoto_2N_trajectories}
\end{figure*}

\begin{figure*}[ht!]
  \begin{overpic}[width=\textwidth]{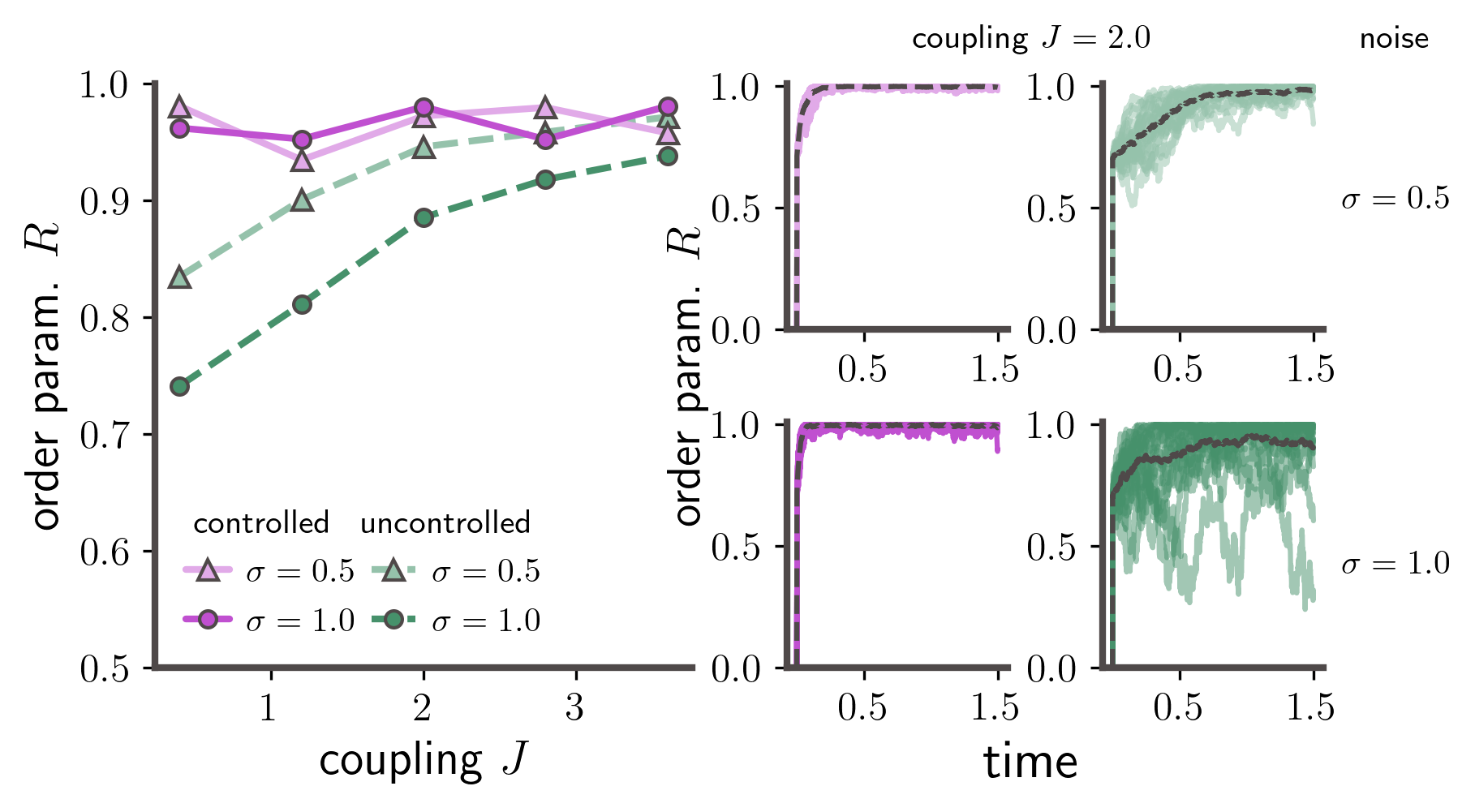}
\put(2,50){\color{black}{\Large\textbf{ a.}}}
\put(44,50){\color{black}{\Large\textbf{ b.}}}
\end{overpic} 
  \caption{ \textbf{Synchronisation control a network of two coupled Kuramoto phase oscillators for different coupling strengths.}  \textbf{(a.)} Time averaged phase-coherence order parameter ($R$) of controlled (\emph{purple}) and uncontrolled (\emph{green}) networks under different noise conditions ($\sigma=0.5$ - \emph{triangles}, $\sigma=1.0$ - \emph{circles}).  The proposed method (\DPF) effectively synchronises the controlled oscillators already for vanishing coupling strength between them. 
   \textbf{(b.)} Evolution of Kuramoto order parameter $R$ for networks with coupling $J=2.0$ and two noise conditions ($\sigma=0.5$ - (upper), $\sigma=1.0$ - (lower))
   for controlled (\emph{purple}) and uncontrolled (\emph{green}) oscillators. The control induces fast transition to synchrony (R=1), while the identical uncontrolled oscillators either synchronise slower (for low noise), or become only partially synchronised (for strong noise). Individual lines indicate evolution of the order parameter in $20$ realisations of the network starting from same initial conditions and from a single computation of the required controls for each setting (where relevant). Dotted black lines denote the mean over the $20$ realisations.
   \footnote{Further parameters: particle number $N=2000$, inducing point number: $M=80$, natural frequencies: $\omega_i = 0$, initial condition: $\theta_i \sim \mathcal{N}(3, 0.5^2)$ , $T=1.5$.} }  
   \label{fig:kuramoto_2N_syst}
\end{figure*}

\begin{figure*}[ht!]
\begin{overpic}[width=\textwidth]{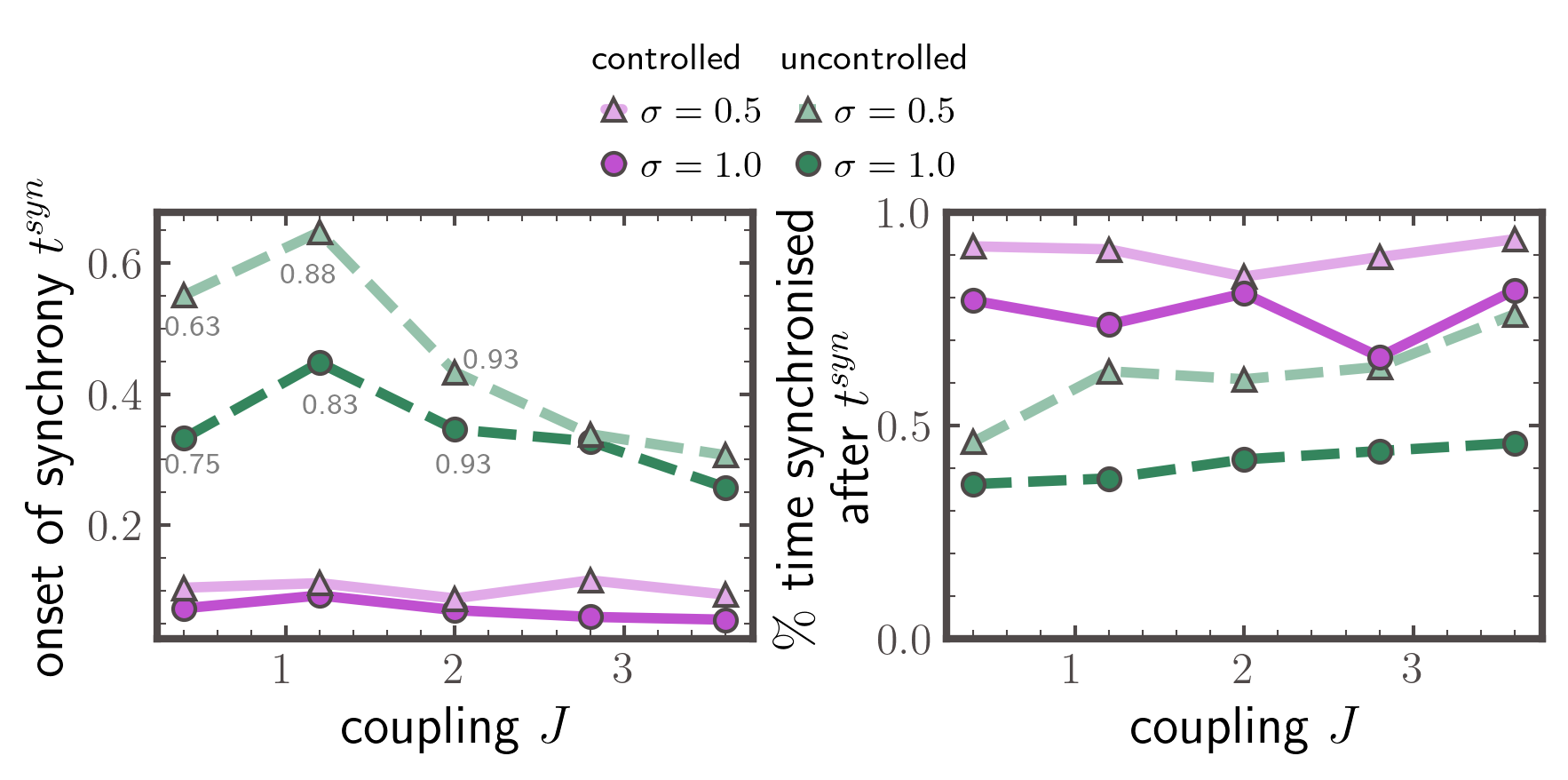}
\put(5,35){\color{black}{\Large\textbf{ a.}}}
\put(52.5,35){\color{black}{\Large\textbf{ b.}}}
\end{overpic}
  \caption[two_kuramotosys_onset]{ \textbf{Onset of synchronisation and percentage of time in synchronised state after synchrony onset reveal the effectiveness of deterministic particle flow control to induce robust synchronisation on a network of $K=2$ oscillators .}  \textbf{(a.)} Onset of synchronisation for controlled (\emph{purple}) and uncontrolled (\emph{green}) networks quantified as the first time $t^{syn}$ the phase-coherence order parameter exceeds $R \geq 0.99$ and remains over that value for duration $\tau^s=20 \times dt = 0.02$ time units. Grey annotations denote the percentage of the examined networks that reached the synchronous state for duration $\tau^s$. Absence of annotation indicates that all examined networks reached synchrony.
   \textbf{(b.)} Percentage of simulation time the networks spontaneously spent in synchronised state ($R \geq 0.99$) after synchrony onset $t^{syn}$. Both figures consider two different noise conditions ($\sigma=0.5$ - \emph{triangles}, $\sigma=1.0$ - \emph{circles}). 
   \footnote{Further parameters: particle number $N=2000$, inducing point number: $M=80$, $T=1.5$. For each noise condition dots denote average over $3$ control computations with different initial conditions with $20$ controlled trajectories for each ($60$ total controlled trajectories for each point). } }
   \label{fig:kuramoto_2N_onset}
\end{figure*}

\begin{figure}[ht!]
 \begin{overpic}[width=0.5\textwidth]{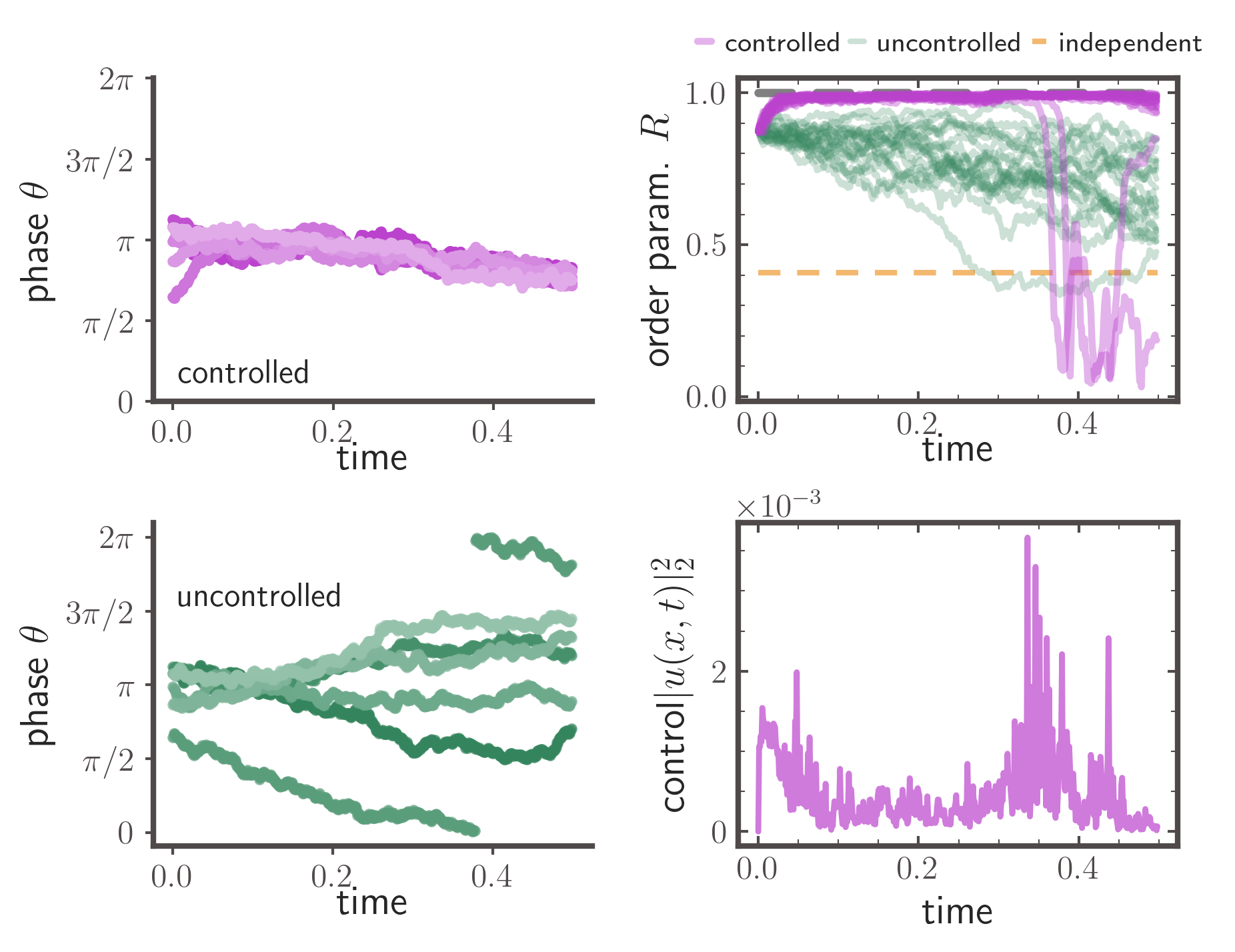}
\put(2,70){\color{black}{\large\textbf{ a.}}}
\put(47,70){\color{black}{\large\textbf{ c.}}}
\put(2,32){\color{black}{\large\textbf{ b.}}}
\put(47,32){\color{black}{\large\textbf{ d.}}}
\end{overpic} 
  \caption{ \textbf{Synchronisation control of a finite-size network of $K=6$ interacting Kuramoto phase oscillators.}  \textbf{(a.)} Evolution of phases $\theta$ of a controlled network, and  \textbf{(b.) } an identical uncontrolled network. The phases of the oscillators quickly synchronise when controlled by \DPF ~and remain synchronised throughout the entire simulation interval $[0,\,T=0.5]$. In the absence of control the phases of the oscillators become increasingly incoherent. \textbf{(c.)} Evolution of Kuramoto order parameter capturing phase coherence $R$ for the controlled (\emph{purple}) oscillator network indicates a rapid transition to complete synchrony ($R=1$) for all $20$ realisations (individual purple lines). The order parameter of identical uncontrolled networks fluctuates strongly indicating partial incoherence (\emph{green}). The orange line denotes the expected long time average value of the order parameter if the oscillators were non-interacting considering finite size scaling effects. For visual clarity, the grey line marks the level of $R=1$ indicating a completely synchronous state. \textbf{(d.)} Control energy spent on all $K=6$ oscillators for a single control realisation.\footnote{Further parameters: coupling strength $J=1.$, noise amplitude $\sigma = 1$, particle number $N=3000$, inducing point number $M=300$, discretisation time step $dt=10^{-3}$, $T=0.5$. }   } 
  \label{fig:kuramoto_instance}
\end{figure}

\begin{figure*}[ht!]
  \begin{overpic}[width=\textwidth]{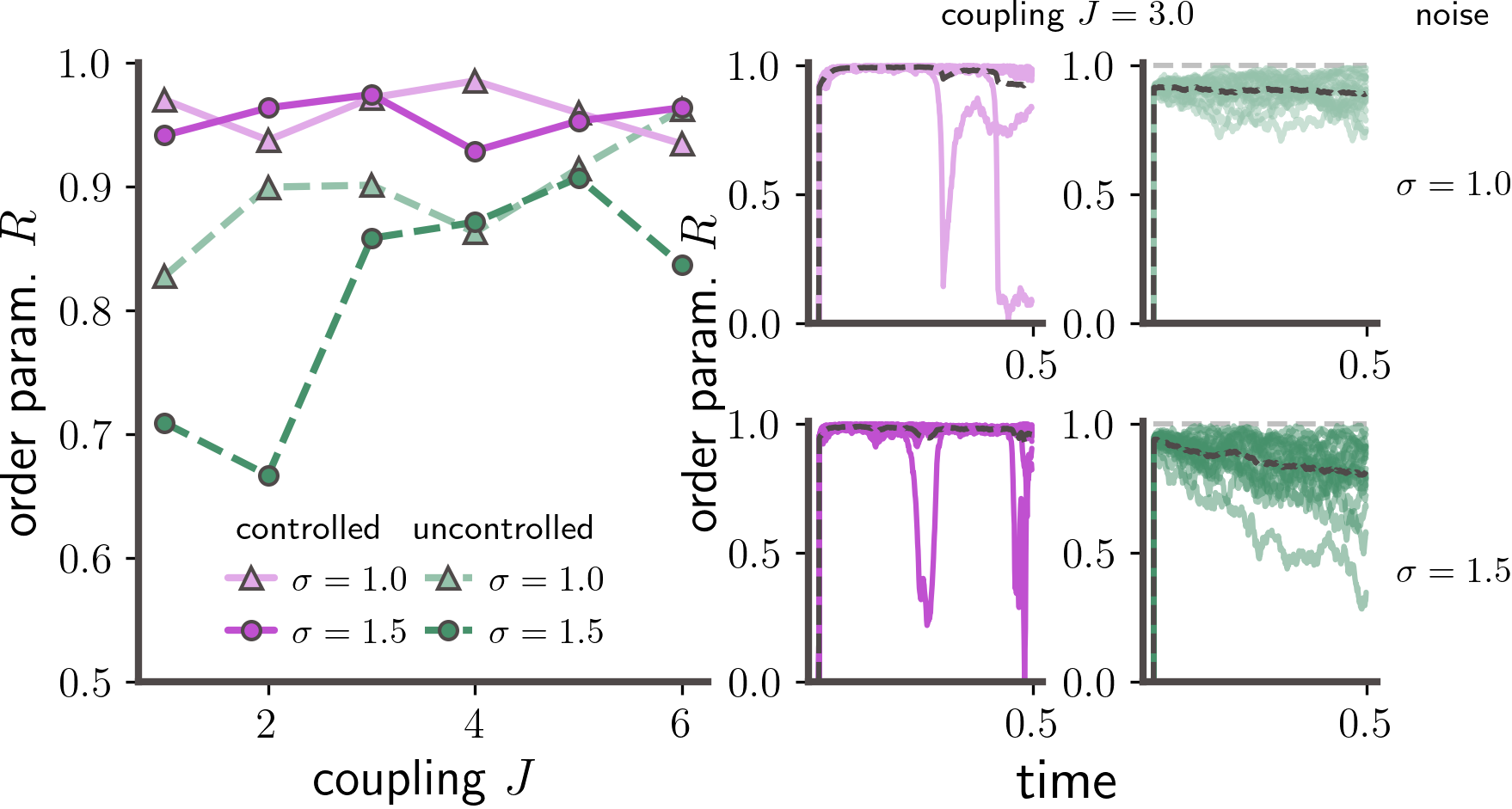}
\put(-1,50){\color{black}{\Large\textbf{ a.}}}
\put(44.5,50){\color{black}{\Large\textbf{ b.}}}
\end{overpic}
  \caption{ \textbf{Synchronisation control of a network of six heterogeneous Kuramoto phase oscillators for different coupling strengths.}  \textbf{(a.)} Time averaged phase-coherence order parameter ($R$) of controlled (\emph{purple}) and uncontrolled (\emph{green}) networks under different noise conditions ($\sigma=1.0$ - \emph{triangles}, $\sigma=1.5$ - \emph{circles}).  The proposed method (\DPF) effectively synchronises the controlled oscillators already for weak coupling. 
   \textbf{(b.)} Evolution of Kuramoto order parameter $R$ for networks with coupling $J=3.0$ and two noise conditions ($\sigma=1.0$ - (upper), $\sigma=1.5$ - (lower))
   for controlled (\emph{purple}) and uncontrolled (\emph{green}) oscillators. The control induces fast transition to synchrony ($R=1$), while the identical uncontrolled oscillators either synchronise slower (for low noise), or become only partially synchronised (for strong noise). Individual lines indicate evolution of the order parameter in $20$ realisations of the network starting from same initial conditions and from a single computation of the required controls for each setting (where relevant). Dotted black lines denote the mean over the $20$ realisations.
   \footnote{Further parameters: particle number $N=3000$, inducing point number: $M=300$, natural frequencies: $\omega_i = 0$, initial condition: $\theta_i \sim \mathcal{N}(3, 0.5^2)$, natural frequencies: $\omega_i \sim \mathcal{N}(0, 1)$, $T$=0.5. Each point in (a.) denotes the mean over $20$ realisations of a single control computation.} }  
   \label{fig:kuramoto_6N_syst}
\end{figure*}

To further demonstrate the generalisability of the proposed framework, we considered a system where the constraints do not explicitly penalise exact regions of the state space, but rather pertain the \emph{collective state} of a system of interacting stochastic units. Specifically, we applied our method for synchronising finite size networks of stochastic Kuramoto phase oscillators (see Supplementary Information for details). We performed systematic studies on a prototypical network of two interacting oscillators (Fig.~\ref{fig:kuramoto_2N_trajectories} , Fig.~\ref{fig:kuramoto_2N_syst}, and Fig.~\ref{fig:kuramoto_2N_onset}), and a network of $K=6$ (Fig.~\ref{fig:kuramoto_instance} and Fig.~\ref{fig:kuramoto_6N_syst}) heterogeneous oscillators with all-to-all uniform coupling (Supplementary Information).

To accommodate synchronisation control with our framework, we considered a time constrained setting where we applied \DPF ~for an interval $[0,\,T]$ (time units). An alternative 'online' approach described in the Supplementary Information alternates computation of controls within small time intervals with simulation of controlled trajectories over those intervals. 
 
We implemented the synchronisation constraint as a path constraint that promotes synchrony without any further requirement for the terminal state, and characterised the level of synchronisation in terms of the Kuramoto order parameter for phase coherence $R(\bm{\theta},t)$ (Supplementary Information)
\begin{equation} \label{eq:order_parameter_main}
    R(t) = \frac{1}{K} | \sum^K_{j=1} e^{i \theta_j (t)} | .  
 \end{equation}
To that end, we employed the following path constraint that promotes order parameter values closer to $1$, i.e. closer to a synchronous state
    \[\mathcal{U}(\bm{\theta}_t) =  \beta\left( 1- R(\bm{\theta}_t,t) \right) dt, \] with $\bm{\theta}_t \in \mathbb{R}^K$ denoting the vector of oscillator phases, and $\beta \in \mathbb{R}$ a scaling constant.

For all considered networks of interacting Kuramoto oscillators we applied interventions $u_{i}(\bm{\theta}_t,t)$ on the phases ${\bm{\theta} = \{\theta_{i}\}^K_{i=1}}$ of all network nodes. Further considerations of optimally selecting a subset of nodes to control were out of the scope of the current article.

For a prototypical network of $K=2$ interacting oscillators with weak coupling ($J=1.2$), 
\DPF ~induced rapid synchrony, whereas the uncontrolled oscillators became progressively incoherent as indicated both by observing their phases (Figure~\ref{fig:kuramoto_2N_trajectories}a.) and the transient values of their phase-coherent order parameter $R$ (Figure~\ref{fig:kuramoto_2N_trajectories}b.). \DPF ~provided fairly strong control inputs at the beginning of the simulation to fully align the phases of the oscillators (Figure~\ref{fig:kuramoto_2N_trajectories}c.), and subsequently delivered only moderate controls to maintain synchrony counteracting the effect of noise.

 \DPF ~successfully induced synchrony for different levels of coupling and under two different noise conditions ${\sigma = \{0.5, 1.0\}}$ (Figure~\ref{fig:kuramoto_2N_syst}a.). For strongly coupled oscillators where the uncontrolled network progressively synchronises solely due to the interactions, \DPF ~synchronised the oscillators faster - indicated by earlier reaching the order parameter ${R=1}$ value compared to their uncontrolled counterparts - and prevented spontaneous desynchronsation events that occurred in the uncontrolled networks especially in the presence of strong noise (Figure~\ref{fig:kuramoto_2N_syst}b.). 

To quantify these effects further, we analysed the onset of synchronisation $t^{syn}$ and the percentage of time spent in the synchronised state of the examined controlled and uncontrolled networks for increasing coupling $J$. 

For each network realisation, we defined the onset of synchronisation $t^{syn}$ as the first time point when the phase-coherence order parameter exceeded the value $R \geq 0.99$ and remained above that value for a minimal duration of $\tau^s=20 \times dt=0.02$ time units. For uncontrolled networks with weak couplings we considered only the subset of realisations that reached the synchronous state (indicated as a ratio by the grey annotations in Figure~\ref{fig:kuramoto_2N_onset}a.).
To quantify the robustness of synchronisation in each network, we estimated the percentage of time the network remained synchronised after the synchronisation onset by counting the time points when the order parameter \emph{spontaneously} exceeded the $R=0.99$ threshold after $t^{syn}$. 

For both noise conditions and independent of coupling strength $J$,  networks controlled by \DPF ~reached the synchronous state considerably faster than their uncontrolled counterparts (Figure~\ref{fig:kuramoto_2N_onset}a.), and consistently remained synchronised for the entire simulation, as mediated by the percentage of time spent in the synchronised state after the synchronisation onset $t^{syn}$. More precisely, while a subset of uncontrolled weakly coupled networks synchronised for at least $\tau^s=20$ time units, as expected due to the presence of noise they failed to remain in that synchronised state as indicated in Figure~\ref{fig:kuramoto_2N_onset}b. For stronger couplings desynchronisation was less pronounced, yet still more frequent than in their controlled counterparts.

For $K=6$ interacting heterogeneous oscillators and for network charachteristics (coupling strength $J = 1. $  and noise amplitude $\sigma = 1.0$) that render the uncontrolled system only partially synchronisable (Figure ~\ref{fig:kuramoto_instance} b.(lower)), state feedback control delivered by \DPF ~successfully drove the oscillators to a fully synchronised state (Figure ~\ref{fig:kuramoto_instance} b.(upper)). As indicated by the evolution of the Kuramoto order parameter for each of the $20$ realisations shown in Figure ~\ref{fig:kuramoto_instance} b., our framework not only delivered sufficient controls to rapidly synchronise the oscillators, but also provided the necessary interventions to maintain the phase synchronisation (Figure ~\ref{fig:kuramoto_instance} c.). In fact, as indicated by the non-fluctuating order parameter $R$  for most realisations, only $2$ network instances underwent spontaneous noise-induced desynchronisations late in the simulations, which nevertheless were partially recovered.

Similar to the smaller network, the six oscillator network was successfully synchronised for a range of coupling strengths (Figure~\ref{fig:kuramoto_6N_syst}a.) and for both noise conditions. Compared to the identical uncontrolled networks, the networks controlled by \DPF ~exhibited consistently larger order parameter values. Although some networks underwent spontaneous desynchronisations, these were quickly efficiently recovered (Figure~\ref{fig:kuramoto_6N_syst}b.).


\section*{Discussion}


In this work, we introduced a novel methodological framework for identifying optimal dynamical interventions for constraining diffusive systems. 
Distinctively from previous work~\cite{kappen2005linear,kappen2016adaptive,wells2015control,nusken2021solving} that devises optimal control protocols by employing iterative optimisation procedures, here, we obtained the required interventions in a \emph{deterministic} and \emph{non-iterative} way. 
In particular, we showed that splitting the time-resolved constraining information into retrospective and prospective parts, allows for a representation of the optimal interventions in terms of the difference of logarithmic gradients (scores) of two forward probability flows. By introducing statistical estimators for the logarithmic gradients of the empirical probability densities, and by  employing novel advances for deterministic evolution of sample based probability flows~\cite{Maoutsa_2020,reich2013nonparametric}, we proposed
an efficient,  non-parametric approximation of the optimal controls.

We demonstrated the feasibility and potential of our framework on a battery of diverse biologically inspired systems and challenging settings of increasing complexity and dimensionality. More precisely, we employed the proposed deterministic particle flow control (\DPF) to induce switches between stable states on multi-stable systems (Section~\ref{res:cell}), to devise artificial selection protocols on phenotypic landscapes by implementing constraints for co-varying phenotypes (Section~\ref{res:evol}), and to induce synchronisation on networks of stochastic phase oscillators.



We compared our approach against the recently proposed Path integral cross entropy method~\cite{kappen2016adaptive}, which approximates time dependent controls by an iterative optimisation based on stochastic path sampling.  Our results suggest that our \emph{one-shot, deterministic} framework is on par with the iterative path integral cross entropy method in terms of control efficiency, and more precise and accurate in terms of deviation from terminal target states.

\if False 
 
On the other hand, the {\em Feynman--Kac theorem} allows the solution of the linear PDE (Eq.~\eqref{backward_equation}) to be expressed in terms of an integral over
stochastic paths from the uncontrolled system, with appropriately reweighted paths to take the constraints into account. 
The {\em path integral cross entropy method} improves over this basic Monte--Carlo approach~\cite{kappen2016adaptive,zhang2014applications} by employing importance sampling. Stochastic paths, generated from an SDE with a non--optimal control, 
are properly reweighted, leading to an iterative optimisation of the control.
\fi
Moreover, through the nonparametric estimation of the logarithmic gradients, our method is inherently model agnostic, being able, in principle, to approximate the logarithmic gradient of arbitrary distributions, and therefore estimate external interventions of arbitrary functional form. In turn, the path integral cross entropy method requires an a priori selection of the number of basis functions employed in the approximation, implicitly constraining the class of control functions that may be implemented. In fact, inappropriate choice of basis function parameters results in numerical instabilities in the form of non converging matrix inversions and diverging trajectories during the optimisation, before the algorithm converges to an optimum. We found that this effect could not be mitigated by adjusting the learning rate or other parameters to smaller values. The trade off, however, pertains the nature of our approximation, which as a kernel method, is more expensive compared to basis functions during the actual evaluation of the identified controls.

In this work, we propagated probability densities by employing recent advances for solving Fokker--Planck equations in terms of deterministic particle dynamics.
However, in principle, any particle filtering algorithm employing stochastic particle dynamics may be combined with the logarithmic gradient density estimator
to obtain a numerical approximation of the time-reversed drift of Eq.~\eqref{eq:backward_drift2} at each time step.
Numerical experiments of such a method showed, that the stochastic fluctuations
of the particles lead to fairly noisy control estimates over time. 
Hence, here, taking advantage of the fact that the {\em deterministic sampling} framework of ~\cite{Maoutsa_2020} already employs  
the logarithmic gradient estimator as a building block, we integrated 
 this method into the computation of the optimal interventions.

Although the representation of constrained densities with particles is computationally more efficient compared to solutions of discretised PDEs, control representations for regions of the state space where the particles do not provide sufficient evidence for the underlying density, will be inaccurate. While the probability density functions for the regions of the state space unoccupied by particles are expected to have vanishing values, with inadequate number of particles, the estimated logarithmic gradient of the related densities might be inaccurate in those regions. Yet, due to the deterministic nature of our approach, our framework provides better representation of the underlying densities compared to a pure stochastic path sampling.

We expressed the optimal interventions as a time and space dependent perturbation of the uncontrolled dynamics that are obtained from the logarithmic gradient of a backward PDE. This formulation results from linearising the Hamilton-Jacobi-Bellman equation (Supplementary Information). An equivalent formulation for optimal drift adjustment for constraining diffusion processes has been derived in the field of statistical mechanics by applying the Doob’s $h$-transform~\cite{orland2011generating,mazzolo2017constrained,szavits2015inequivalence,majumdar2015effective,chetrite2015variational}. This formulation agrees with the one employed here in the absence of path constraints. In that case, the optimal interventions may be obtained from the logarithmic gradient of the solution of the backward Kolmogorov equation. However, the solution of the backward Kolmogorov equation is known in closed form only for trivial systems, may be cumbersome to obtain for general multidimensional systems, while the ensuing computation of the logarithmic gradient of the solution may run into numerical instabilities . 
Here, by representing the densities with particles, and by directly estimating the logarithmic gradients of the particle density, we provided an efficient and feasible solution for obtaining the necessary drift adjustments.

The present work proposes a method for controlling stochastic systems by manipulating their dynamical variables. However, often such a scenario might be insufficient. Interventions in biological systems may have limited access to the the system's state variables, or only system parameters might be accessible for control.
In these settings, an extension of the path integral control framework that additionally considers the system parameters is required. Existing methods that consider parameter optimisation are limited to properly function only in weak noise settings~\cite{wells2015control} due to large deviations arguments employed in their derivation.


%

We successfully employed the proposed method for synchronisation control of networks of coupled stochastic Kuramoto oscillators. However, when path constraints are required, the \DPF ~solves an optimal transport problem at every time step to implement the path constraints as a deterministic particle resampling. This computation employs the Earth Mover's distance (EMD) for an ensemble of $N$ particles, which scales rather unfavorably for increasing particle number as $\mathcal{O}(N^3 \log N)$. Here, for the networked systems that required large number of particles, we employed an alternative solution for computing the EMD, the network simplex solver~\cite{bonneel2011displacement, flamary2021pot}, which has computational complexity that scales as $\mathcal{O}(N^2 )$, providing thereby a considerable speedup in the calculations. Yet,  we still consider this necessary particle shifting a computational bottleneck for our framework.
 Future developments will focus on incorporating this step into the forward particle dynamics and thus will scale more favourably with particle number, enabling thereby the control of systems of higher dimensionality.

Moreover, considering the topic of network synchronisation, we considered out of the scope of the present paper to explore the possibility of controlling only a subset of network nodes. Previous attempts to solve the same problem in a stochastic setting have considered only the synchronisation of two coupled identical Kuramoto oscillators~\cite{li2015optimal}, while, here, we considered networks up to six oscillators with differing natural frequencies. Insights from network control theory for nonlinear systems coupled with the proposed framework may provide a more control-energy efficient approach for network synchronisation.  Additionally, further systematic studies that will explore various network topologies, coupling schemes, and intrinsic parameter heterogeneities, will provide additional insight on the properties of our method to induce robust synchronisation to networks of interacting stochastic phase oscillators.




The proposed framework is further relevant for several computational or applied settings, where marginal densities of constrained diffusive systems are required, i.e. for state estimation of such systems between two successive state observations, or for computing the transition probabilities in extreme event calculations.
In those settings, only the constrained path distribution $q_t(x)$ is required instead of the precise dynamical interventions. Although not explicitly demonstrated here, \DPF ~is also applicable for computing averages over constrained densities or functionals over constrained paths. Since the reverse time sampled flow $\tilde{q}_t(x)$ already provides a good representation of the constrained density, averages over functions evaluated on the paths of $q_t(x)$ already provide accurate estimates of the computed quantities.

More broadly, the field of simulation based inference~\cite{brehmer2021simulation,cranmer2020frontier} has developed dramatically in the past years due to expanding computational capacities offered by current computing devises able to simulate models considered demanding in the past. These methods perform inference for models with intractable likelihoods (like sparsely observed stochastic systems) requiring frameworks for accurate and efficient simulation of detailed dynamical models that provide dynamical trajectories of candidate models to perform inference. Thereby, our approach, by providing direct samplings of dynamically constrained diffusion processes will enable efficient inference of such systems.

We employed the proposed framework on a prototypical scenario of devising artificial selection protocols for molecular phenotypes inspired by ~\cite{nourmohammad2021optimal}.
The nascent field of continuous culturing~\cite{alon2019introduction,gresham2014enduring} for studying adaptive evolution has created a growing demand for devising efficient and precise stochastic control frameworks that may be integrated in advanced open-source continuous culturing platforms like eVOLVER~\cite{evolver} to manipulate and control cell culture growth and phenotyping. These platforms enable real time monitoring of cell cultures and administer \emph{exact} custom perturbations in the form of selection pressures, or by adjusting the constant nutrient feed input to the culture. By providing accurate interventions that implement arbitrary state constraints the proposed \DPF ~control is suitable to be integrated in such a platform.

The clearest advantage of our framework is its non-iterative, non-parametric, and deterministic nature, providing thereby computational advantages compared to existing control methods~\cite{kappen2016adaptive,hairer2009sampling}. Moreover, the proposed formulation for the optimal interventions generalises beyond particle systems and may be easily implemented by neural networks.



\bibliography{scibib}

\begin{thebibliography}{89}
\providecommand{\natexlab}[1]{#1}
\providecommand{\url}[1]{\texttt{#1}}
\expandafter\ifx\csname urlstyle\endcsname\relax
  \providecommand{\doi}[1]{doi: #1}\else
  \providecommand{\doi}{doi: \begingroup \urlstyle{rm}\Url}\fi

\bibitem[Raser and O'shea(2005)]{raser2005noise}
Jonathan~M Raser and Erin~K O'shea.
\newblock \href{https://doi.org/10.1126/science.1105891}{Noise in gene
  expression: origins, consequences, and control}.
\newblock \emph{Science}, 309\penalty0 (5743):\penalty0 2010--2013, 2005.

\bibitem[Swain et~al.(2002)Swain, Elowitz, and Siggia]{swain2002intrinsic}
Peter~S Swain, Michael~B Elowitz, and Eric~D Siggia.
\newblock \href{https://doi.org/10.1073/pnas.162041399}{Intrinsic and extrinsic
  contributions to stochasticity in gene expression}.
\newblock \emph{Proceedings of the National Academy of Sciences}, 99\penalty0
  (20):\penalty0 12795--12800, 2002.

\bibitem[Hasty et~al.(2000)Hasty, Pradines, Dolnik, and
  Collins]{hasty2000noise}
Jeff Hasty, Joel Pradines, Milos Dolnik, and James~J Collins.
\newblock \href{https://doi.org/10.1073/pnas.040411297}{Noise-based switches
  and amplifiers for gene expression}.
\newblock \emph{Proceedings of the National Academy of Sciences}, 97\penalty0
  (5):\penalty0 2075--2080, 2000.

\bibitem[Kepler and Elston(2001)]{kepler2001stochasticity}
Thomas~B Kepler and Timothy~C Elston.
\newblock \href{https://doi.org/10.1016/S0006-3495(01)75949-8}{Stochasticity in
  transcriptional regulation: origins, consequences, and mathematical
  representations}.
\newblock \emph{Biophysical journal}, 81\penalty0 (6):\penalty0 3116--3136,
  2001.

\bibitem[Garc{\'\i}a-Ojalvo and Sancho(2012)]{garcia2012noise}
Jordi Garc{\'\i}a-Ojalvo and Jos{\'e} Sancho.
\newblock \emph{\href{https://doi.org/10.1007/978-1-4612-1536-3}{Noise in
  spatially extended systems}}.
\newblock Springer Science \& Business Media, 2012.

\bibitem[Eldar and Elowitz(2010)]{eldar2010functional}
Avigdor Eldar and Michael~B Elowitz.
\newblock \href{https://doi.org/10.1038/nature09326}{Functional roles for noise
  in genetic circuits}.
\newblock \emph{Nature}, 467\penalty0 (7312):\penalty0 167--173, 2010.

\bibitem[Simpson et~al.(2009)Simpson, Cox, Allen, McCollum, Dar, Karig, and
  Cooke]{simpson2009noise}
Michael~L Simpson, Chris~D Cox, Michael~S Allen, James~M McCollum, Roy~D Dar,
  David~K Karig, and John~F Cooke.
\newblock \href{https://doi.org/10.1002/wnan.22}{Noise in biological circuits}.
\newblock \emph{Wiley Interdisciplinary Reviews: Nanomedicine and
  Nanobiotechnology}, 1\penalty0 (2):\penalty0 214--225, 2009.

\bibitem[Blake et~al.(2003)Blake, K{\ae}rn, Cantor, and
  Collins]{blake2003noise}
William~J Blake, Mads K{\ae}rn, Charles~R Cantor, and James~J Collins.
\newblock \href{https://doi.org/10.1038/nature01546}{Noise in eukaryotic gene
  expression}.
\newblock \emph{Nature}, 422\penalty0 (6932):\penalty0 633--637, 2003.

\bibitem[Horsthemke(1984)]{horsthemke1984noise}
Werner Horsthemke.
\newblock \href{https://doi.org/10.1007/3-540-36852-3}{Noise induced
  transitions}.
\newblock In \emph{Non-equilibrium dynamics in chemical systems}, pages
  150--160. Springer, 1984.

\bibitem[Zhou et~al.(2005)Zhou, Chen, and Aihara]{zhou2005molecular}
Tianshou Zhou, Luonan Chen, and Kazuyuki Aihara.
\newblock \href{https://doi.org/10.1103/PhysRevLett.95.178103}{Molecular
  communication through stochastic synchronization induced by extracellular
  fluctuations}.
\newblock \emph{Physical Review Letters}, 95\penalty0 (17):\penalty0 178103,
  2005.

\bibitem[Del~Vecchio et~al.(2016)Del~Vecchio, Dy, and Qian]{del2016control}
Domitilla Del~Vecchio, Aaron~J Dy, and Yili Qian.
\newblock \href{https://doi.org/10.1098/rsif.2016.0380}{Control theory meets
  synthetic biology}.
\newblock \emph{Journal of The Royal Society Interface}, 13\penalty0
  (120):\penalty0 20160380, 2016.

\bibitem[Nguyen et~al.(2021)Nguyen, Pease, and Kueh]{nguyen2021scalable}
Phuc Nguyen, Nicholas~A Pease, and Hao~Yuan Kueh.
\newblock \href{https://doi.org/10.1098/rsif.2021.0109}{Scalable control of
  developmental timetables by epigenetic switching networks}.
\newblock \emph{Journal of the Royal Society Interface}, 18\penalty0
  (180):\penalty0 20210109, 2021.

\bibitem[Sivak and Crooks(2012)]{sivak2012thermodynamic}
David~A Sivak and Gavin~E Crooks.
\newblock \href{https://arxiv.org/pdf/1201.4166.pdf}{Thermodynamic metrics and
  optimal paths}.
\newblock \emph{Physical Review Letters}, 108\penalty0 (19):\penalty0 190602,
  2012.

\bibitem[Zulkowski et~al.(2012)Zulkowski, Sivak, Crooks, and
  DeWeese]{zulkowski2012geometry}
Patrick~R Zulkowski, David~A Sivak, Gavin~E Crooks, and Michael~R DeWeese.
\newblock \href{http://dx.doi.org/10.1103/PhysRevE.86.041148}{Geometry of
  thermodynamic control}.
\newblock \emph{Physical Review E}, 86\penalty0 (4):\penalty0 041148, 2012.

\bibitem[Gomez-Marin et~al.(2008)Gomez-Marin, Schmiedl, and
  Seifert]{gomez2008optimal}
Alex Gomez-Marin, Tim Schmiedl, and Udo Seifert.
\newblock \href{https://doi.org/10.1063/1.2948948}{Optimal protocols for
  minimal work processes in underdamped stochastic thermodynamics}.
\newblock \emph{The Journal of chemical physics}, 129\penalty0 (2):\penalty0
  024114, 2008.

\bibitem[Rabitz et~al.(2000)Rabitz, de~Vivie-Riedle, Motzkus, and
  Kompa]{rabitz2000whither}
Herschel Rabitz, Regina de~Vivie-Riedle, Marcus Motzkus, and Karl Kompa.
\newblock \href{https://doi.org/10.1126/science.288.5467.824}{Whither the
  future of controlling quantum phenomena?}
\newblock \emph{Science}, 288\penalty0 (5467):\penalty0 824--828, 2000.

\bibitem[Pechen and Tannor(2011)]{pechen2011there}
Alexander~N Pechen and David~J Tannor.
\newblock \href{https://doi.org/10.1103/PhysRevLett.106.120402}{Are there traps
  in quantum control landscapes?}
\newblock \emph{Physical Review Letters}, 106\penalty0 (12):\penalty0 120402,
  2011.

\bibitem[Hendrix and Jarzynski(2001)]{hendrix2001fast}
DA~Hendrix and C~Jarzynski.
\newblock A “fast growth” method of computing free energy differences.
\newblock \emph{The Journal of Chemical Physics}, 114\penalty0 (14):\penalty0
  5974--5981, 2001.

\bibitem[Shirts et~al.(2003)Shirts, Bair, Hooker, and
  Pande]{shirts2003equilibrium}
Michael~R Shirts, Eric Bair, Giles Hooker, and Vijay~S Pande.
\newblock \href{https://doi.org/10.1103/PhysRevLett.91.140601}{Equilibrium free
  energies from nonequilibrium measurements using maximum-likelihood methods}.
\newblock \emph{Physical Review Letters}, 91\penalty0 (14):\penalty0 140601,
  2003.

\bibitem[Schmiedl and Seifert(2007)]{schmiedl2007optimal}
Tim Schmiedl and Udo Seifert.
\newblock Optimal finite-time processes in stochastic thermodynamics.
\newblock \emph{Physical Review Letters}, 98\penalty0 (10):\penalty0 108301,
  2007.

\bibitem[Hartmann and Sch{\"u}tte(2012)]{hartmann2012efficient}
Carsten Hartmann and Christof Sch{\"u}tte.
\newblock \href{https://page.mi.fu-berlin.de/chartman/rare.pdf}{Efficient rare
  event simulation by optimal nonequilibrium forcing}.
\newblock \emph{Journal of Statistical Mechanics: Theory and Experiment},
  2012\penalty0 (11):\penalty0 P11004, 2012.

\bibitem[Chetrite and Touchette(2015)]{chetrite2015variational}
Rapha{\"e}l Chetrite and Hugo Touchette.
\newblock \href{https://arxiv.org/pdf/1506.05291.pdf}{Variational and optimal
  control representations of conditioned and driven processes}.
\newblock \emph{Journal of Statistical Mechanics: Theory and Experiment},
  2015\penalty0 (12):\penalty0 P12001, 2015.

\bibitem[Kim and Mehta(2020)]{kim2020optimal}
Jin~Won Kim and Prashant~G Mehta.
\newblock \href{https://arxiv.org/pdf/1904.01710.pdf}{An optimal control
  derivation of nonlinear smoothing equations}.
\newblock In \emph{Proceedings of the Workshop on Dynamics, Optimization and
  Computation held in honor of the 60th birthday of Michael Dellnitz}, pages
  295--311. Springer, 2020.

\bibitem[Casadiego et~al.(2018)Casadiego, Maoutsa, and
  Timme]{casadiego2018inferring}
Jose Casadiego, Dimitra Maoutsa, and Marc Timme.
\newblock \href{https://doi.org/10.1103/PhysRevLett.121.054101}{Inferring
  network connectivity from event timing patterns}.
\newblock \emph{Physical review letters}, 121\penalty0 (5):\penalty0 054101,
  2018.

\bibitem[Todorov(2009)]{todorov2009efficient}
Emanuel Todorov.
\newblock \href{https://www.pnas.org/content/106/28/11478}{Efficient
  computation of optimal actions}.
\newblock \emph{Proceedings of the National Academy of Sciences of the United
  States of America}, 106\penalty0 (28):\penalty0 11478--11483, 2009.

\bibitem[Kappen(2005{\natexlab{a}})]{kappen2005linear}
Hilbert~J Kappen.
\newblock \href{https://arxiv.org/abs/physics/0411119}{Linear theory for
  control of nonlinear stochastic systems}.
\newblock \emph{Physical Review Letters}, 95\penalty0 (20):\penalty0 200201,
  2005{\natexlab{a}}.

\bibitem[Wells et~al.(2015)Wells, Kath, and Motter]{wells2015control}
Daniel~K Wells, William~L Kath, and Adilson~E Motter.
\newblock \href{https://doi.org/10.1103/PhysRevX.5.031036}{Control of
  stochastic and induced switching in biophysical networks}.
\newblock \emph{Physical Review X}, 5\penalty0 (3):\penalty0 031036, 2015.

\bibitem[Nourmohammad and Eksin(2021)]{nourmohammad2021optimal}
Armita Nourmohammad and Ceyhun Eksin.
\newblock
  \href{https://journals.aps.org/prx/abstract/10.1103/PhysRevX.11.011044}{Optimal
  evolutionary control for artificial selection on molecular phenotypes}.
\newblock \emph{Physical Review X}, 11\penalty0 (1):\penalty0 011044, 2021.

\bibitem[Zhong et~al.(2020)Zhong, Wong, Ravikumar, Arzumanyan, Khalil, and
  Liu]{evolver}
Ziwei Zhong, Brandon~G Wong, Arjun Ravikumar, Garri~A Arzumanyan, Ahmad~S
  Khalil, and Chang~C Liu.
\newblock \href{https://pubs.acs.org/doi/10.1021/acssynbio.0c00135}{Automated
  continuous evolution of proteins in-vivo}.
\newblock \emph{ACS synthetic biology}, 9\penalty0 (6):\penalty0 1270--1276,
  2020.

\bibitem[Kao et~al.(2021)Kao, Sadabadi, and Hennequin]{kao2021optimal}
Ta-Chu Kao, Mahdieh~S Sadabadi, and Guillaume Hennequin.
\newblock \href{https://doi.org/10.1016/j.neuron.2021.03.009}{Optimal
  anticipatory control as a theory of motor preparation: a thalamo-cortical
  circuit model}.
\newblock \emph{Neuron}, 109\penalty0 (9):\penalty0 1567--1581, 2021.

\bibitem[Iolov et~al.(2014)Iolov, Ditlevsen, and Longtin]{iolov2014stochastic}
Alexandre Iolov, Susanne Ditlevsen, and Andr{\'e} Longtin.
\newblock \href{https://doi.org/10.1088/1741-2560/11/4/046004}{Stochastic
  optimal control of single neuron spike trains}.
\newblock \emph{Journal of Neural Engineering}, 11\penalty0 (4):\penalty0
  046004, 2014.

\bibitem[Scott(2004)]{scott2004optimal}
Stephen~H Scott.
\newblock \href{https://doi.org/10.1038/nrn1427}{Optimal feedback control and
  the neural basis of volitional motor control}.
\newblock \emph{Nature Reviews Neuroscience}, 5\penalty0 (7):\penalty0
  532--545, 2004.

\bibitem[Bernton et~al.(2019)Bernton, Heng, Doucet, and Jacob]{bernton2019schr}
Espen Bernton, Jeremy Heng, Arnaud Doucet, and Pierre~E Jacob.
\newblock \href{https://arxiv.org/abs/1912.13170}{Schr{\"o}dinger Bridge
  Samplers}.
\newblock \emph{arXiv preprint arXiv:1912.13170}, 2019.

\bibitem[Vargas et~al.(2021)Vargas, Thodoroff, Lawrence, and
  Lamacraft]{vargas2021solving}
Francisco Vargas, Pierre Thodoroff, Neil~D Lawrence, and Austen Lamacraft.
\newblock \href{https://arxiv.org/abs/2106.02081}{Solving Schr{\"o}dinger
  Bridges via Maximum Likelihood}.
\newblock \emph{arXiv preprint arXiv:2106.02081}, 2021.

\bibitem[Exarchos and Theodorou(2018)]{exarchos2018stochastic}
Ioannis Exarchos and Evangelos~A Theodorou.
\newblock \href{https://doi.org/10.1016/j.automatica.2017.09.004}{Stochastic
  optimal control via forward and backward stochastic differential equations
  and importance sampling}.
\newblock \emph{Automatica}, 87:\penalty0 159--165, 2018.

\bibitem[Song et~al.(2020)Song, Sohl-Dickstein, Kingma, Kumar, Ermon, and
  Poole]{song2020score}
Yang Song, Jascha Sohl-Dickstein, Diederik~P Kingma, Abhishek Kumar, Stefano
  Ermon, and Ben Poole.
\newblock \href{https://arxiv.org/abs/2011.13456}{Score-based generative
  modeling through stochastic differential equations}.
\newblock \emph{arXiv preprint arXiv:2011.13456}, 2020.

\bibitem[Bellman(1956)]{bellman1956dynamic}
Richard Bellman.
\newblock \href{https://www.pnas.org/content/42/10/767}{Dynamic programming and
  Lagrange multipliers}.
\newblock \emph{Proceedings of the National Academy of Sciences of the United
  States of America}, 42\penalty0 (10):\penalty0 767, 1956.

\bibitem[Garcke and Kr{\"o}ner(2017)]{garcke2017suboptimal}
Jochen Garcke and Axel Kr{\"o}ner.
\newblock \href{https://doi.org/10.1007/s10915-016-0240-7}{Suboptimal feedback
  control of PDEs by solving HJB equations on adaptive sparse grids}.
\newblock \emph{Journal of Scientific Computing}, 70\penalty0 (1):\penalty0
  1--28, 2017.

\bibitem[Annunziato and Borz{\`\i}(2013)]{annunziato2013fokker}
Mario Annunziato and Alfio Borz{\`\i}.
\newblock \href{https://doi.org/10.1016/j.cam.2012.06.019}{A Fokker--Planck
  control framework for multidimensional stochastic processes}.
\newblock \emph{Journal of Computational and Applied Mathematics}, 237\penalty0
  (1):\penalty0 487--507, 2013.

\bibitem[Horowitz et~al.(2014)Horowitz, Damle, and Burdick]{horowitz2014linear}
Matanya~B Horowitz, Anil Damle, and Joel~W Burdick.
\newblock \href{https://doi.org/10.1109/CDC.2014.7040310}{Linear Hamilton
  Jacobi Bellman equations in high dimensions}.
\newblock In \emph{53rd IEEE Conference on Decision and Control}, pages
  5880--5887. IEEE, 2014.

\bibitem[Kappen(2005{\natexlab{b}})]{kappen2005path}
Hilbert~J Kappen.
\newblock \href{https://doi.org/10.1088/1742-5468/2005/11/P11011}{Path
  integrals and symmetry breaking for optimal control theory}.
\newblock \emph{Journal of statistical mechanics: theory and experiment},
  2005\penalty0 (11):\penalty0 P11011, 2005{\natexlab{b}}.

\bibitem[Van Den~Broek et~al.(2008)Van Den~Broek, Wiegerinck, and
  Kappen]{van2008graphical}
Bart Van Den~Broek, Wim Wiegerinck, and Bert Kappen.
\newblock \href{https://doi.org/10.1613/jair.2473}{Graphical model inference in
  optimal control of stochastic multi-agent systems}.
\newblock \emph{Journal of Artificial Intelligence Research}, 32:\penalty0
  95--122, 2008.

\bibitem[Rawlik et~al.(2013)Rawlik, Toussaint, and Vijayakumar]{rawlik2013path}
Konrad Rawlik, Marc Toussaint, and Sethu Vijayakumar.
\newblock \href{https://arxiv.org/abs/1208.2523}{Path integral control by
  reproducing kernel Hilbert space embedding}.
\newblock In \emph{Twenty-Third International Joint Conference on Artificial
  Intelligence}, 2013.

\bibitem[Kappen and Ruiz(2016)]{kappen2016adaptive}
Hilbert~Johan Kappen and Hans~Christian Ruiz.
\newblock
  \href{https://link.springer.com/content/pdf/10.1007/s10955-016-1446-7.pdf}{Adaptive
  importance sampling for control and inference}.
\newblock \emph{Journal of Statistical Physics}, 162\penalty0 (5):\penalty0
  1244--1266, 2016.

\bibitem[Zhang et~al.(2014)Zhang, Wang, Hartmann, Weber, and
  Schuette]{zhang2014applications}
Wei Zhang, Han Wang, Carsten Hartmann, Marcus Weber, and Christof Schuette.
\newblock \href{https://doi.org/10.1137/14096493X}{Applications of the
  cross-entropy method to importance sampling and optimal control of
  diffusions}.
\newblock \emph{SIAM Journal on Scientific Computing}, 36\penalty0
  (6):\penalty0 A2654--A2672, 2014.

\bibitem[Theodorou et~al.(2011)Theodorou, Stulp, Buchli, and
  Schaal]{theodorou2011iterative}
Evangelos Theodorou, Freek Stulp, Jonas Buchli, and Stefan Schaal.
\newblock \href{https://doi.org/10.3182/20110828-6-IT-1002.02249}{An iterative
  path integral stochastic optimal control approach for learning robotic
  tasks}.
\newblock \emph{IFAC Proceedings Volumes}, 44\penalty0 (1):\penalty0
  11594--11601, 2011.

\bibitem[Thijssen and Kappen(2015)]{thijssen2015path}
Sep Thijssen and HJ~Kappen.
\newblock \href{https://doi.org/10.1103/PhysRevE.91.032104}{Path integral
  control and state-dependent feedback}.
\newblock \emph{Physical Review E}, 91\penalty0 (3):\penalty0 032104, 2015.

\bibitem[Todorov(2008)]{todorov2008general}
Emanuel Todorov.
\newblock \href{https://doi.org/10.1109/CDC.2008.4739438}{General duality
  between optimal control and estimation}.
\newblock In \emph{2008 47th IEEE Conference on Decision and Control}, pages
  4286--4292. IEEE, 2008.

\bibitem[Kappen et~al.(2012)Kappen, G{\'o}mez, and Opper]{kappen2012optimal}
Hilbert~J Kappen, Vicen{\c{c}} G{\'o}mez, and Manfred Opper.
\newblock \href{https://arxiv.org/abs/0901.0633}{Optimal control as a graphical
  model inference problem}.
\newblock \emph{Machine Learning}, 87\penalty0 (2):\penalty0 159--182, 2012.

\bibitem[Levine(2018)]{levine2018reinforcement}
Sergey Levine.
\newblock \href{https://arxiv.org/abs/1805.00909}{Reinforcement learning and
  control as probabilistic inference: Tutorial and review}.
\newblock \emph{arXiv preprint arXiv:1805.00909}, 2018.

\bibitem[Attias(2003)]{attias2003planning}
Hagai Attias.
\newblock \href{http://proceedings.mlr.press/r4/attias03a.html}{Planning by
  probabilistic inference}.
\newblock In \emph{International Workshop on Artificial Intelligence and
  Statistics}, pages 9--16. PMLR, 2003.

\bibitem[Maoutsa et~al.(2020)Maoutsa, Reich, and Opper]{Maoutsa_2020}
Dimitra Maoutsa, Sebastian Reich, and Manfred Opper.
\newblock \href{https://arxiv.org/abs/2006.00702}{Interacting particle
  solutions of Fokker--Planck equations through gradient--log--density
  estimation}.
\newblock \emph{Entropy}, 22\penalty0 (8):\penalty0 802, 2020.

\bibitem[Anderson(1982)]{anderson1982reverse}
Brian~DO Anderson.
\newblock \href{https://doi.org/10.1016/0304-4149(82)90051-5}{Reverse-time
  diffusion equation models}.
\newblock \emph{Stochastic Processes and their Applications}, 12\penalty0
  (3):\penalty0 313--326, 1982.

\bibitem[Gillespie and Petzold(2003)]{gillespie2003improved}
Daniel~T Gillespie and Linda~R Petzold.
\newblock \href{https://doi.org/10.1063/1.1613254}{Improved leap-size selection
  for accelerated stochastic simulation}.
\newblock \emph{The Journal of Chemical Physics}, 119\penalty0 (16):\penalty0
  8229--8234, 2003.

\bibitem[Saarinen et~al.(2008)Saarinen, Linne, and
  Yli-Harja]{saarinen2008stochastic}
Antti Saarinen, Marja-Leena Linne, and Olli Yli-Harja.
\newblock \href{https://doi.org/10.1371/journal.pcbi.1000004}{Stochastic
  differential equation model for cerebellar granule cell excitability}.
\newblock \emph{PLoS Computational Biology}, 4\penalty0 (2):\penalty0 e1000004,
  2008.

\bibitem[Lande et~al.(2003)Lande, Engen, Saether, et~al.]{lande2003stochastic}
Russell Lande, Steinar Engen, Bernt-Erik Saether, et~al.
\newblock \emph{Stochastic population dynamics in ecology and conservation}.
\newblock Oxford University Press on Demand, 2003.

\bibitem[Takahata et~al.(1975)Takahata, Ishii, and Matsuda]{takahata1975effect}
Naoyuki Takahata, Kazushige Ishii, and Hirotsugu Matsuda.
\newblock \href{https://doi.org/10.1073/pnas.72.11.4541}{Effect of temporal
  fluctuation of selection coefficient on gene frequency in a population}.
\newblock \emph{Proceedings of the National Academy of Sciences}, 72\penalty0
  (11):\penalty0 4541--4545, 1975.

\bibitem[Todorov(2005)]{todorov2005stochastic}
Emanuel Todorov.
\newblock \href{https://doi.org/10.1162/0899766053491887}{Stochastic optimal
  control and estimation methods adapted to the noise characteristics of the
  sensorimotor system}.
\newblock \emph{Neural computation}, 17\penalty0 (5):\penalty0 1084--1108,
  2005.

\bibitem[Todorov(2004)]{todorov2004optimality}
Emanuel Todorov.
\newblock \href{https://www.nature.com/articles/nn1309}{Optimality principles
  in sensorimotor control}.
\newblock \emph{Nature neuroscience}, 7\penalty0 (9):\penalty0 907--915, 2004.

\bibitem[Todorov(2007)]{todorov2007linearly}
Emanuel Todorov.
\newblock
  \href{https://dl.acm.org/doi/10.5555/2976456.2976628}{Linearly-solvable
  Markov decision problems}.
\newblock In \emph{Advances in neural information processing systems}, pages
  1369--1376, 2007.

\bibitem[Fleming(1977)]{fleming1977exit}
Wendell~H Fleming.
\newblock \href{https://link.springer.com/article/10.1007/BF01442148}{Exit
  probabilities and optimal stochastic control}.
\newblock \emph{Applied Mathematics and Optimization}, 4\penalty0 (1):\penalty0
  329--346, 1977.

\bibitem[Macris and Marino(2020)]{macris2020solving}
Nicolas Macris and Raffaele Marino.
\newblock \href{https://arxiv.org/abs/2012.07747}{Solving non-linear Kolmogorov
  equations in large dimensions by using deep learning: a numerical comparison
  of discretization schemes}.
\newblock \emph{arXiv preprint arXiv:2012.07747}, 2020.

\bibitem[Li et~al.(2020)Li, Kovachki, Azizzadenesheli, Liu, Bhattacharya,
  Stuart, and Anandkumar]{li2020fourier}
Zongyi Li, Nikola Kovachki, Kamyar Azizzadenesheli, Burigede Liu, Kaushik
  Bhattacharya, Andrew Stuart, and Anima Anandkumar.
\newblock \href{https://arxiv.org/abs/2010.08895}{Fourier neural operator for
  parametric partial differential equations}.
\newblock \emph{International Conference on Learning Representations}, 2020.

\bibitem[Majumdar and Orland(2015)]{majumdar2015effective}
Satya~N Majumdar and Henri Orland.
\newblock \href{https://doi.org/10.1088/1742-5468/2015/06/P06039}{Effective
  Langevin equations for constrained stochastic processes}.
\newblock \emph{Journal of Statistical Mechanics: Theory and Experiment},
  2015\penalty0 (6):\penalty0 P06039, 2015.

\bibitem[Note1()]{Note1}
Note1.
\newblock In the second equality, we considered that for small $\delta t$ the
  commutator of the two operators ${\protect \cal {L}}_f$ and $U(x,t)$ is
  negligible.

\bibitem[Reich(2013)]{reich2013nonparametric}
Sebastian Reich.
\newblock \href{https://arxiv.org/abs/1210.0375}{A nonparametric ensemble
  transform method for Bayesian inference}.
\newblock \emph{SIAM Journal on Scientific Computing}, 35\penalty0
  (4):\penalty0 A2013--A2024, 2013.

\bibitem[De~Bortoli et~al.(2021)De~Bortoli, Thornton, Heng, and
  Doucet]{de2021diffusion}
Valentin De~Bortoli, James Thornton, Jeremy Heng, and Arnaud Doucet.
\newblock \href{https://arxiv.org/abs/2106.01357}{Diffusion Schr$\backslash$"
  odinger Bridge with Applications to Score-Based Generative Modeling}.
\newblock \emph{arXiv preprint arXiv:2106.01357}, 2021.

\bibitem[Gardiner(2009)]{gardiner2009stochastic}
Crispin~W.. Gardiner.
\newblock \emph{Stochastic Methods: A Handbook for the Natural and Social
  Sciences}.
\newblock Springer., 2009.

\bibitem[Villani(2009)]{villani2009optimal}
C{\'e}dric Villani.
\newblock \emph{Optimal transport: old and new}, volume 338.
\newblock Springer, 2009.

\bibitem[Huang et~al.(2007)Huang, Guo, May, and Enver]{huang2007bifurcation}
Sui Huang, Yan-Ping Guo, Gillian May, and Tariq Enver.
\newblock \href{https://doi.org/10.1016/j.ydbio.2007.02.036}{Bifurcation
  dynamics in lineage-commitment in bipotent progenitor cells}.
\newblock \emph{Developmental Biology}, 305\penalty0 (2):\penalty0 695--713,
  2007.

\bibitem[Lande(1976)]{lande1976natural}
Russell Lande.
\newblock \href{https://doi.org/10.1111/j.1558-5646.1976.tb00911.x}{Natural
  selection and random genetic drift in phenotypic evolution}.
\newblock \emph{Evolution}, pages 314--334, 1976.

\bibitem[Fisher(1930)]{fisher1958genetical}
Ronald~Aylmer Fisher.
\newblock \emph{The genetical theory of natural selection}.
\newblock The Clarendon Press, 1930.

\bibitem[Kingsolver et~al.(2001)Kingsolver, Hoekstra, Hoekstra, Berrigan,
  Vignieri, Hill, Hoang, Gibert, and Beerli]{kingsolver2001strength}
Joel~G Kingsolver, Hopi~E Hoekstra, Jon~M Hoekstra, David Berrigan, Sacha~N
  Vignieri, CE~Hill, Anhthu Hoang, Patricia Gibert, and Peter Beerli.
\newblock \href{https://doi.org/10.1086/319193}{The strength of phenotypic
  selection in natural populations}.
\newblock \emph{The American Naturalist}, 157\penalty0 (3):\penalty0 245--261,
  2001.

\bibitem[Barton et~al.(2007)Barton, Briggs, Eisen, Goldstein, and
  Patel]{barton2007evolution}
N.H. Barton, D.E. Briggs, J.A. Eisen, D.B. Goldstein, and N.H. Patel.
\newblock \emph{Evolution}.
\newblock Cold Spring Harbor Laboratory Series. Cold Spring Harbor Laboratory
  Press, 2007.
\newblock ISBN 9780879696849.
\newblock URL \url{https://books.google.de/books?id=mMDFQ32oMI8C}.

\bibitem[Whitlock(1995)]{whitlock1995variance}
Michael~C Whitlock.
\newblock
  \href{https://doi.org/10.1111/j.1558-5646.1995.tb02237.x}{Variance-induced
  peak shifts}.
\newblock \emph{Evolution}, 49\penalty0 (2):\penalty0 252--259, 1995.

\bibitem[Nourmohammad et~al.(2013)Nourmohammad, Schiffels, and
  L{\"a}ssig]{nourmohammad2013evolution}
Armita Nourmohammad, Stephan Schiffels, and Michael L{\"a}ssig.
\newblock \href{https://doi.org/10.1088/1742-5468/2013/01/P01012}{Evolution of
  molecular phenotypes under stabilizing selection}.
\newblock \emph{Journal of Statistical Mechanics: Theory and Experiment},
  2013\penalty0 (01):\penalty0 P01012, 2013.

\bibitem[Held et~al.(2014)Held, Nourmohammad, and L{\"a}ssig]{held2014adaptive}
Torsten Held, Armita Nourmohammad, and Michael L{\"a}ssig.
\newblock \href{https://doi.org/10.1088/1742-5468/2014/09/P09029}{Adaptive
  evolution of molecular phenotypes}.
\newblock \emph{Journal of Statistical Mechanics: Theory and Experiment},
  2014\penalty0 (9):\penalty0 P09029, 2014.

\bibitem[N{\"u}sken and Richter(2021)]{nusken2021solving}
Nikolas N{\"u}sken and Lorenz Richter.
\newblock \href{https://doi.org/10.1007/s42985-021-00102-x}{Solving
  high-dimensional Hamilton--Jacobi--Bellman PDEs using neural networks:
  perspectives from the theory of controlled diffusions and measures on path
  space}.
\newblock \emph{Partial Differential Equations and Applications}, 2\penalty0
  (4):\penalty0 1--48, 2021.

\bibitem[Orland(2011)]{orland2011generating}
Henri Orland.
\newblock \href{https://doi.org/10.1063/1.5000423}{Generating transition paths
  by Langevin bridges}.
\newblock \emph{The Journal of chemical physics}, 134\penalty0 (17):\penalty0
  174114, 2011.

\bibitem[Mazzolo(2017)]{mazzolo2017constrained}
Alain Mazzolo.
\newblock \href{https://doi.org/10.1088/1742-5468/aa4f15}{Constrained Brownian
  processes and constrained Brownian bridges}.
\newblock \emph{Journal of Statistical Mechanics: Theory and Experiment},
  2017\penalty0 (2):\penalty0 023203, 2017.

\bibitem[Szavits-Nossan and Evans(2015)]{szavits2015inequivalence}
Juraj Szavits-Nossan and Martin~R Evans.
\newblock \href{https://doi.org/10.1088/1742-5468/2015/12/P12008}{Inequivalence
  of nonequilibrium path ensembles: the example of stochastic bridges}.
\newblock \emph{Journal of Statistical Mechanics: Theory and Experiment},
  2015\penalty0 (12):\penalty0 P12008, 2015.

\bibitem[Bonneel et~al.(2011)Bonneel, Van De~Panne, Paris, and
  Heidrich]{bonneel2011displacement}
Nicolas Bonneel, Michiel Van De~Panne, Sylvain Paris, and Wolfgang Heidrich.
\newblock Displacement interpolation using lagrangian mass transport.
\newblock In \emph{Proceedings of the 2011 SIGGRAPH Asia Conference}, pages
  1--12, 2011.

\bibitem[Flamary et~al.(2021)Flamary, Courty, Gramfort, Alaya, Boisbunon,
  Chambon, Chapel, Corenflos, Fatras, Fournier, et~al.]{flamary2021pot}
R{\'e}mi Flamary, Nicolas Courty, Alexandre Gramfort, Mokhtar~Zahdi Alaya,
  Aur{\'e}lie Boisbunon, Stanislas Chambon, Laetitia Chapel, Adrien Corenflos,
  Kilian Fatras, Nemo Fournier, et~al.
\newblock \href{http://jmlr.org/papers/v22/20-451.html}{POT: Python optimal
  transport}.
\newblock \emph{Journal of Machine Learning Research}, 22\penalty0
  (78):\penalty0 1--8, 2021.

\bibitem[Li et~al.(2015)Li, Zhang, and Wang]{li2015optimal}
Jr-Shin Li, Wei Zhang, and Shuo Wang.
\newblock \href{https://doi.org/10.1016/j.ifacol.2015.11.015}{Optimal Control
  and Stochastic Synchronization of Phase Oscillators}.
\newblock \emph{IFAC-PapersOnLine}, 48\penalty0 (18):\penalty0 83--88, 2015.

\bibitem[Brehmer(2021)]{brehmer2021simulation}
Johann Brehmer.
\newblock \href{https://doi.org/10.1038/s42254-021-00305-6}{Simulation-based
  inference in particle physics}.
\newblock \emph{Nature Reviews Physics}, 3\penalty0 (5):\penalty0 305--305,
  2021.

\bibitem[Cranmer et~al.(2020)Cranmer, Brehmer, and Louppe]{cranmer2020frontier}
Kyle Cranmer, Johann Brehmer, and Gilles Louppe.
\newblock \href{https://doi.org/10.1073/pnas.1912789117}{The frontier of
  simulation-based inference}.
\newblock \emph{Proceedings of the National Academy of Sciences of the United
  States of America}, 117\penalty0 (48):\penalty0 30055--30062, 2020.

\bibitem[Alon(2019)]{alon2019introduction}
Uri Alon.
\newblock \emph{An introduction to systems biology: design principles of
  biological circuits}.
\newblock CRC press, 2019.

\bibitem[Gresham and Dunham(2014)]{gresham2014enduring}
David Gresham and Maitreya~J Dunham.
\newblock \href{https://doi.org/10.1016/j.ygeno.2014.09.015}{The enduring
  utility of continuous culturing in experimental evolution}.
\newblock \emph{Genomics}, 104\penalty0 (6):\penalty0 399--405, 2014.

\bibitem[Hairer et~al.(2009)Hairer, Stuart, and Voss]{hairer2009sampling}
Martin Hairer, Andrew~M Stuart, and Jochen Voss.
\newblock \href{https://www.jstor.org/stable/29779386}{Sampling conditioned
  diffusions}.
\newblock \emph{Trends in stochastic analysis}, 353:\penalty0 159--186, 2009.

\end{thebibliography}

\section*{Acknowledgments}
We thank Sebastian Reich for insightful discussions during the early development of this work. This research has been partially funded by Deutsche Forschungsgemeinschaft (DFG)-SFB1294/ 1-318763901.\\



\end{document}